\documentclass[10pt,conference]{IEEEtran}
\IEEEoverridecommandlockouts
\usepackage[letterpaper,top=0.7in,bottom=1in,left=0.7in,right=0.7in,
            columnsep=0.1in]{geometry}

\usepackage{cite}
\usepackage{amsmath,amssymb,amsfonts}
\usepackage{graphicx}
\usepackage{textcomp}
\usepackage{xcolor}
\usepackage[hyphens]{url}
\usepackage{fancyhdr}

\newcommand{\hpcayear}{2027}

\newcommand{\hpcasubmissionnumber}{1612}
\title{\Name: A Unified Programming Framework for Retargetable PIM Compilation}

\def\hpcacameraready{}

\newcommand\hpcaauthors{Niansong Zhang, Shiran Guo\textsuperscript{*}, and Zhiru Zhang%
  \thanks{\textsuperscript{*}Shiran Guo is with the Hong Kong University of
  Science and Technology (HKUST). This work was done during his in-person research
  internship at Cornell University.}}
\newcommand\hpcaaffiliation{Cornell University}
\newcommand\hpcaemail{\{nz264, sg2682, zhiruz\}@cornell.edu}

\usepackage{listings}
\usepackage{mathtools}
\usepackage{booktabs}
\usepackage{multirow}
\usepackage{xspace}
\usepackage{makecell}
\usepackage{caption}
\usepackage{subcaption}
\usepackage{adjustbox}
\usepackage{mdframed}
\usepackage{siunitx}
\usepackage{algorithm}
\usepackage{enumitem}
\usepackage{algpseudocode}

\usepackage{tikz}
\usetikzlibrary{arrows.meta,positioning,fit,backgrounds,calc,shapes.geometric}

\definecolor{pimred}{RGB}{228,80,70}
\definecolor{pimgray}{RGB}{190,190,190}
\definecolor{pimfill}{RGB}{247,247,247}

\tikzset{
  pimbox/.style   = {rounded corners=2pt, draw=black, line width=0.9pt,
                     inner sep=3pt, align=center, font=\scriptsize},
  pimuser/.style  = {pimbox, draw=pimred, text=pimred},
  pimstage/.style = {pimbox, fill=pimfill},
  pimgroup/.style = {rounded corners=3pt, draw=pimgray, line width=0.7pt,
                     dashed, inner sep=4pt},
  pimpath/.style  = {-{Latex[length=1.6mm,width=1.4mm]}, line width=0.8pt},
  pimdash/.style  = {pimpath, dashed, line width=0.7pt},
  pimlbl/.style   = {font=\scriptsize, inner sep=1.5pt},
  pimtiny/.style  = {font=\tiny, inner sep=1pt},
  pimpanel/.style = {font=\footnotesize},
}

\usepackage{pifont}
\makeatletter
\def\subsubsection{\@startsection{subsubsection}{3}{\z@}
  {0ex plus 0.1ex minus 0.1ex}{0ex}{\normalfont\normalsize\itshape}}
\def\paragraph{\@startsection{paragraph}{4}{\z@}
  {0ex plus 0.1ex minus 0.1ex}{0ex}{\normalfont\normalsize\itshape}}
\makeatother

\usepackage{hyperref}
\hypersetup{hidelinks,pdfauthor={Niansong Zhang, Shiran Guo, Zhiru Zhang}}
\newcommand{\revise}[1]{{#1}}
\usepackage[T1]{fontenc}
\usepackage{plex-mono}
\usepackage{fvextra}
\input{highlight-style}

\newcommand{\Name}{Tenon\xspace}

\newcommand\code[1]{{\small\ttfamily #1}}

\providecommand{\Description}[1]{}

\newenvironment{acks}{\section*{Acknowledgments}}{}

\author{
  \ifdefined\hpcacameraready
    \IEEEauthorblockN{\hpcaauthors{}}
      \IEEEauthorblockA{
        \hpcaaffiliation{} \\
        \hpcaemail{}
      }
  \else
    \IEEEauthorblockN{\normalsize{HPCA \hpcayear{} Submission
      \textbf{\#\hpcasubmissionnumber{}}} \\
      \IEEEauthorblockA{
        Confidential Draft \\
        Do NOT Distribute!!
      }
    }
  \fi 
}

\fancypagestyle{camerareadyfirstpage}{%
  \fancyhead{}
  
  \fancyhead[C]{
    \ifdefined\aeopen
    \parbox[][12mm][t]{13.5cm}{\hpcayear{} IEEE International Symposium on High-Performance Computer Architecture (HPCA)}    
    \else
      \ifdefined\aereviewed
      \parbox[][12mm][t]{13.5cm}{\hpcayear{} IEEE International Symposium on High-Performance Computer Architecture (HPCA)}
      \else
      \ifdefined\aereproduced
      \parbox[][12mm][t]{13.5cm}{\hpcayear{} IEEE International Symposium on High-Performance Computer Architecture (HPCA)}
      \else
      \parbox[][0mm][t]{13.5cm}{\hpcayear{} IEEE International Symposium on High-Performance Computer Architecture (HPCA)}
    \fi 
    \fi 
    \fi 
    \ifdefined\aeopen 
      \includegraphics[width=12mm,height=12mm]{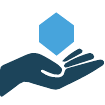}
    \fi 
    \ifdefined\aereviewed
      \includegraphics[width=12mm,height=12mm]{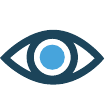}
    \fi 
    \ifdefined\aereproduced
      \includegraphics[width=12mm,height=12mm]{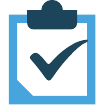}
    \fi
  }
  \fancyfoot[C]{}
}
\begin{document}
\maketitle

\ifdefined\hpcacameraready 
  \thispagestyle{camerareadyfirstpage}
  \pagestyle{empty}
\else
  \thispagestyle{plain}
  \pagestyle{plain}
\fi

\newcommand{\hpcaheight}{0mm}
\ifdefined\eaopen
\renewcommand{\hpcaheight}{12mm}
\fi



\begin{abstract}
Processing-in-memory (PIM) is emerging as a building block of heterogeneous
systems for memory-bound workloads, and commercial prototypes now span near-bank
SIMD engines, bank-level GDDR accelerators, programmable DRAM cores, and
bit-level in-SRAM processors.  Their programming stacks are fragmented: no abstraction or
compiler is shared across vendors, and per-device models range from
instruction sequences to C-like languages. This fragmentation makes PIM solutions difficult to evaluate and adopt. Data layout further complicates the problem: tight compute-memory integration makes data layout a first-order optimization, yet existing PIM compilers lack a unified way to represent and optimize it.
A key observation is that PIM compilation targets can be modeled as nested
grids of compute and storage units connected by legal data-transfer
paths. Work and data are then mapped jointly onto this structure.
We present \Name, a unified programming framework for
retargetable PIM compilation. It decouples workload, compilation target, and cost
specifications, so a workload is expressed once and compiled through a unified
programming framework. \Name generalizes the $\mathbb{F}_2$ linear layout algebra
from GPU compilers to map logical work and data onto these compute and storage
grids. It automatically constructs a minimum-movement data layout given a workload and a target specification.
We evaluate \Name on six PIM backends: five commercial backends and
AttAcc, a state-of-the-art academic HBM-based PIM architecture. Across the
commercial backends, it matches or outperforms vendor,
expert, and compiler baselines, with geometric-mean speedups of up to
$1.20\times$ and individual speedups up to $3.65\times$.
For end-to-end GPT-13B decoding on AttAcc, \Name-compiled workloads
run $1.2$--$1.3\%$ faster than those compiled by DCC, a state-of-the-art PIM
compiler, across all three batch-4 configurations.
Together, these results demonstrate \Name's performance portability
across diverse PIM target families.
\end{abstract}



\section{Introduction}
\label{sec:introduction}

Processing-in-memory (PIM) is moving from
research prototypes toward deployment.  Memory-bound workloads, including
batched large language model (LLM) inference, motivate heterogeneous systems
that assign memory-bound layers to PIM devices
~\cite{park2024attacc,heo2024neupims,seo2024ianus}.  A growing set of
commercial implementations demonstrates this progress, including Samsung
HBM-PIM near-bank SIMD~\cite{kwon2021hbmpim}, SK Hynix AiM bank-level GDDR6
processors~\cite{lee2022aim}, UPMEM programmable DRAM
cores~\cite{gomez2022prim}, and GSI compute-in-SRAM vector processors
~\cite{gwennap2020gemini}.  These targets differ in their instructions, memory
hierarchies, parallelism, and host interfaces.

Their programming stacks remain fragmented
(Table~\ref{tab:pim-stacks}).  Vendor programming models range from
low-level instruction sequences to C-like languages. Existing infrastructure
such as CINM~\cite{singh2024cinm} uses the Multi-Level Intermediate
Representation (MLIR)~\cite{lattner2021mlir} across heterogeneous
compute-in-memory and compute-near-memory targets but retains target-specific
device dialects and lowering.
Evaluating a workload across different PIM targets therefore still requires repeated mapping and optimization effort for each device.

\begin{figure*}[t]
  \centering
  \includegraphics[width=\textwidth]{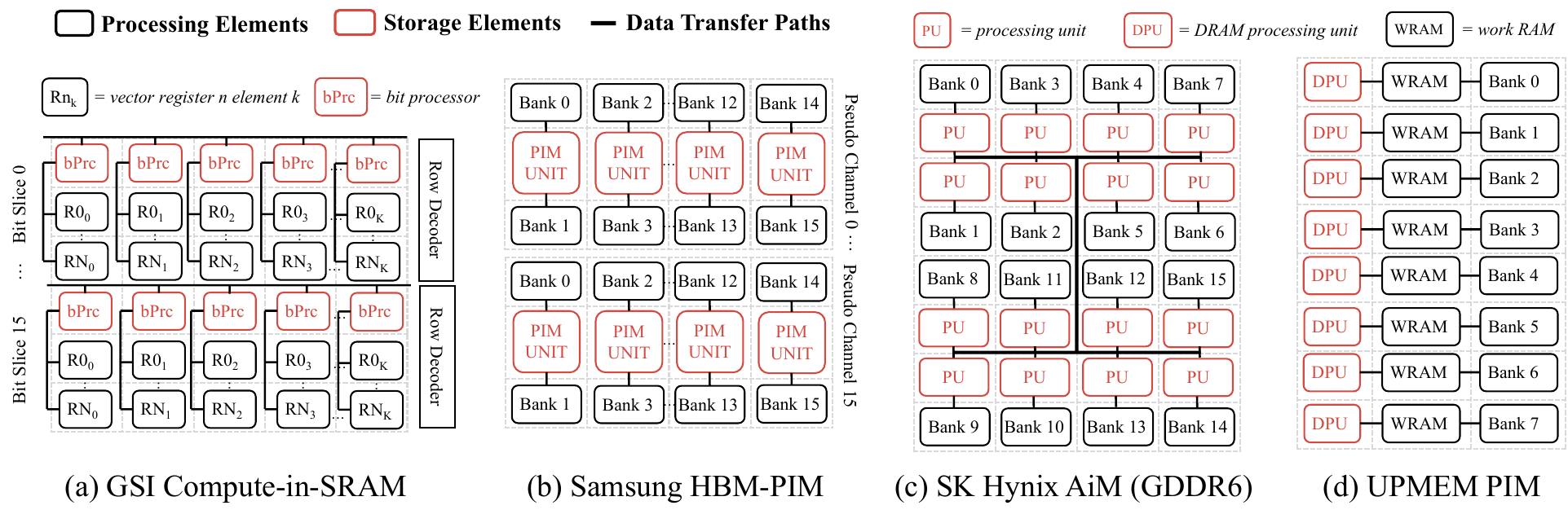}
  \caption{\textbf{Representative commercial PIM architectures} -- GSI couples
  vector operations to SRAM; Samsung and SK Hynix place SIMD units beside DRAM
  banks; and UPMEM pairs a RISC core with each DRAM bank. In every case,
  compute directly reaches only its local memory partition, so non-local
  operands require explicit host-mediated transfers.}
  \label{fig:pim-archs}
\end{figure*}

\begin{table*}[t]
\centering
\footnotesize
\caption{\textbf{Compilation stacks for representative commercial PIM targets}
-- Each vendor defines a distinct program form, toolchain, data-layout and
movement \revise{requirements}; third-party compiler support is uneven.}
\label{tab:pim-stacks}
\setlength{\tabcolsep}{4pt}
\begin{adjustbox}{max width=\textwidth}
\begin{tabular}{@{}lllll@{}}
\toprule
\textbf{Backend} & \textbf{Programmer writes} & \textbf{Vendor toolchain}
        & \textbf{Data layout and movement} & \textbf{Third-party compilers} \\
\midrule
Samsung HBM-PIM
  & PCU op code stream
  & PIMLibrary,
  & Implicit bank-row reads;
  & \revise{---} \\
  &
  & PIMSimulator~\cite{saitpimlibrary,saitpimsimulator}
  & bank/channel layout by hand
  & \\[2pt]

SK Hynix AiM
  & ISR command trace
  & Ramulator2 AiM fork~\cite{luo2024ramulator2},
  & Every command explicit;
  & Expert mappings \\
  & (no imperative program)
  & timing only
  & GB staging for cross-bank
  & (CENT~\cite{gu2025cent}) \\[2pt]

UPMEM
  & C kernel per DPU
  & LLVM-based \texttt{dpu-upmem-}
  & Explicit MRAM/WRAM DMA;
  & PrIM~\cite{gomez2022prim}, SimplePIM~\cite{chen2023simplepim}, \\
  & $+$ host driver
  & \texttt{dpurte-clang}; uPIMulator~\cite{hyun2024upimulator}
  & no DPU-to-DPU path
  & CINM~\cite{singh2024cinm}, ATiM~\cite{shin2025atim} \\[2pt]

GSI Gemini-I
  & Host C $+$ device C
  & GDL/GVML SDK~\cite{zhang2025apu}
  & Explicit DMA versus PIO
  & None \\
  & with GVML calls
  &
  & across L4--L1 and VRs
  & \\[2pt]

GSI Gemini-II
  & Single-file C++ program
  & \texttt{gsi-gtml-g2} SDK~\cite{gsi2026tools}
  & Copy to/from L1;
  & None \\
  & over a tensor library
  &
  & vectors pinned to 64K lanes
  & \\
\bottomrule
\end{tabular}
\end{adjustbox}

\par\vspace{3pt}
\begin{minipage}{\textwidth}
\raggedright
\scriptsize
\textit{Abbreviations:}
PCU, Programmable Computing Unit; ISR, Instruction Set Register;
GB, Global Buffer;
\revise{DPU, Data Processing Unit; MRAM, Main RAM; WRAM, Working RAM;
DMA, Direct Memory Access; PIO, Programmed Input/Output; VR, Vector Register;}
GDL, GSI Device Library; GVML, GSI Vector Math Library.
\end{minipage}

\end{table*}

Optimizing data layout for PIM is a second challenge.  Data layout optimizations that assign elements to banks, replicate operands, or arrange elements within vector registers affect PIM performance by an order of magnitude.  Reported gains include
$6.86\times$ for PIMnast GEMV over its host
SoC~\cite{ibrahim2024pimnast}, $72.8\times$ for OptiPIM over an HBM-PIM
mapping heuristic~\cite{liu2025optipim}, and $18.9\times$ for binary matrix multiplication on GSI compute-in-SRAM~\cite{zhang2025apu}.
A PIM compute unit can directly access only part of the target's storage.  A
data layout therefore determines which unit can execute an operation and what
data movement is required.  Existing PIM compilers encode these decisions with
device-specific address calculations and expert schedules.  They lack a common
representation for these layout decisions across architectures.

General tensor accelerators typically separate compute arrays from distributed memories and connect them through programmable on-chip networks and cache hierarchies. They often include multiple memory levels and distinct load, compute, and store engines. Achieving high performance therefore requires coordinating loop schedules, tiling, spatial dataflows, memory allocation, communication, and pipelining. 
PIM architectures expose a stronger common invariant despite their diversity: computation is tightly coupled with a statically identifiable memory partition. Accessing data outside the partition is typically costly or impossible without host intervention. This coupling reduces computation mapping and data layout to a constrained placement-and-movement problem.

We present \Name, a unified programming framework for
retargetable PIM compilation.
\Name exploits this reduced scheduling dimensionality to make data layout optimization, analytical cost modeling, and compilation retargeting possible across families of PIM targets.
\Name models a PIM compilation target as nested grids of compute
and storage units connected by legal data transfer paths. Compiling a kernel
then requires jointly mapping logical work and data onto this structure
and inserting explicit data transfers.

\Name generalizes the $\mathbb{F}_2$ linear layout algebra from GPU
compilers~\cite{zhou2026linearlayouts} to represent these mappings.  Because
regular PIM axes have sizes that are powers of two, their coordinates are bit
vectors. One map expresses bank interleaving, tile splitting, and operand
replication, while algebraic checks establish coverage and detect conflicts.
Unlike a GPU swizzle, a PIM layout specifies which attached compute unit owns
an operand and therefore which operations and transfers are legal.  \Name
carries this layout contract through packing, lowering, and cost evaluation,
replacing address formulas written separately for each device with one
unified layout and mapping representation.  Its minimum-movement layout construction uses
operand reuse subspaces, workload shape, and target transfer costs to construct
a legal layout with minimum data-movement cost.

We evaluate \Name on six PIM targets: five commercial targets and
AttAcc, a state-of-the-art academic HBM-based PIM
architecture~\cite{park2024attacc}. Across the commercial targets, \Name
matches Samsung's vendor GEMV and
reaches expert parity on physical Gemini-I ($1.013\times$). It outperforms the
best UPMEM, Gemini-II, and AiM baselines by geometric means of
$1.20\times$, $1.19\times$, and $1.10\times$, respectively.
For end-to-end GPT-13B decoding on AttAcc, \Name matches
DCC, a state-of-the-art PIM compiler~\cite{yang2025dcc}, to reported precision
at batch~1 and
\Name-compiled workloads run $1.2$--$1.3\%$ faster than those compiled by DCC
across all three batch-4 configurations.

Executable cost programs support both schedule optimization and composition
with external tools. A separate AttAcc reference-best study finds that \Name
matches the best archived timing on 398 of 500 instances versus DCC's 326;
the attention reference certificates are conditional on a stopping assumption.
\Name's component expressions provide a structured interface for
agent-assisted calibration. In a Gemini-I case study, an agent fits
target-specific operation costs from component microbenchmarks and composes
them to reduce latency-prediction error from $10.67\%$ to $1.29\%$.
Target and cost programs also form a virtual backend for workload-driven
architectural exploration without a simulator or hardware. On AiM, this reveals
a knee at a $4\,\mathrm{KB}$ Global Buffer. Wider rows and larger
buffers enable larger reduction tiles and fewer commands, but further scaling
has diminishing returns once payload movement dominates.

This paper makes four contributions:

\begin{itemize}[left=0pt]
    \item \textbf{A retargetable PIM programming framework and compilation flow.}
          \Name decouples workloads, targets, and cost specification using a unified
          programming framework.  It represents each target as nested compute and storage
          grids with declared operations and transfer paths.

    \item \textbf{Automatic layout construction for PIM spatial mapping.}
          \Name generalizes the $\mathbb{F}_2$ linear layout algebra to represent data
          layouts and regular spatial work mappings across PIM hierarchies.
          Because data layout constrains compute ownership on PIM, these relations
          drive legality checks, packing, lowering, transfers, and cost
          evaluation.  The compiler constructs a minimum-movement legal PIM layout
          for the given workload shape and target hierarchy.

    \item \textbf{Reusable schedule search.}
          Schedule spaces reuse a shared workflow for
          enumeration, legality checking, code generation, and cost evaluation.

    \item \textbf{Evaluation across physical, simulated, and virtual targets.}
          We evaluate five commercial PIM backends and AttAcc. We use
          cost programs for agent-assisted calibration and workload-driven
          architecture exploration, where the target and cost
          specifications act as a virtual backend.
  \end{itemize}


\section{Motivation and Background}
\label{sec:motivation}
\Name compiles workloads to diverse PIM targets that share no instruction set, memory hierarchy, or host interface. A single architectural property makes this retargetability possible. We identify that property next, then show why it makes data layout a central compiler decision and how one algebra expresses that decision across PIM architectures.

\begin{figure*}[t]
  \centering
  \includegraphics[width=0.86\textwidth]{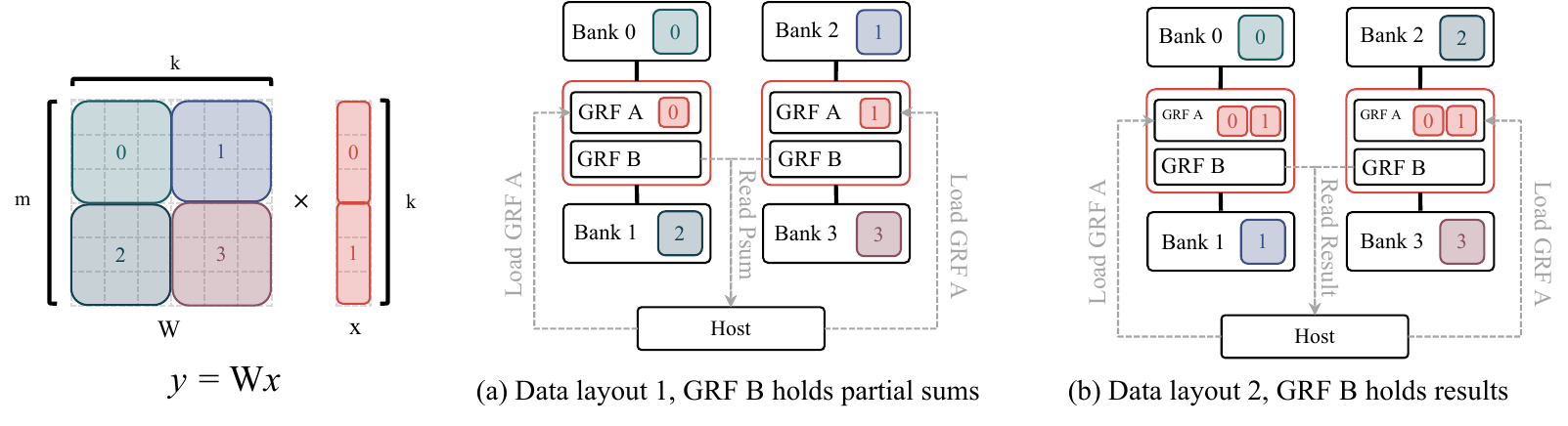}
  \caption{\textbf{Samsung data layouts trade input replication against host
  reduction} -- For matrix-vector multiplication $y=Wx$, each PIM unit owns
  one even/odd bank pair, and the MAC is
  $\texttt{GRF\_B}[m]\mathbin{+}=\texttt{GRF\_A}[n]\times\texttt{BANK}[\mathit{row},\mathit{col}]$,
  so $W$ streams from a bank while $x$ must be resident in \texttt{GRF\_A}.
  Both layouts fill all four banks, so the choice is not about parallelism.
  (a) Cutting $W$ along $k$ partitions $x$, but \revise{leaves} a partial sum in every
  unit and no inter-unit datapath can combine them.  (b) Cutting $W$ along $m$
  lets a local MAC finish each output, at the price of replicating $x$.
  Replication scales with $k$ and partial gathering with $m$, so (b) suits tall
  matrices ($m>k$) and (a) wide ones ($k>m$); (b) also needs enough rows to
  fill the units.}
  \label{fig:samsung-gemv-layout}
  \Description{On the left, the product y = Wx with W drawn as a two-by-two
  grid of colored blocks numbered 0 to 3 and x as two blocks numbered 0 and 1.
  Two panels follow. Each shows two PIM units, each holding a GRF A and a GRF B
  register between an even and an odd DRAM bank, with a host box below. In
  panel (a) the W blocks are assigned so that each unit holds a k-slice of
  every output row, each unit's GRF A holds one block of x, and the host reads
  partial sums back. In panel (b) the W blocks are assigned by output row, each
  unit's GRF A holds both blocks of x, and the host reads finished results
  back.}
\end{figure*}

\subsection{What Makes a Shared PIM Abstraction Possible}
\label{sec:pim-property}

Figure~\ref{fig:pim-archs} and Table~\ref{tab:pim-stacks} summarize the
commercial targets considered in this work: Samsung near-bank SIMD, AiM single- and all-bank
operations, UPMEM multithreaded DRAM cores, and two generations of GSI
compute-in-SRAM vector processors
~\cite{kwon2021hbmpim,lee2022aim,gomez2022prim,zhang2025apu}. Their ISAs,
memory hierarchies, interfaces, and useful schedules have little in common.

One defining property that spans these devices is that compute is attached to a memory
partition, so placement determines which unit can operate on a value. A
near-bank unit, DPU, or SRAM vector array reads only physically wired storage;
no transparent cache or dynamic scheduler decouples address from compute
ownership. On a conventional processor placement is a locality hint. On PIM it
is a legality constraint, and reaching a non-local operand requires an
expensive explicit transfer.

This coupling makes layout a target-specific scheduling decision. On DRAM-PIM,
work mapping and bank layout must agree. Samsung additionally requires a tile
to use both banks attached to each PIM unit
(Section~\ref{sec:layout-motivation}). On GSI, position within a vector register
determines whether movement crosses the horizontal mux network, which costs
roughly an order of magnitude more than an inter-register move
~\cite{zhang2025apu}. Thus, the architecture fixes available arithmetic while
the compiler jointly chooses work mapping and data layout.

The common compiler problem is therefore to partition work, place operands so
each unit can access them, choose local operations, and price required
transfers. \Name represents this structure as nested grids of compute and
storage units connected by legal transfer paths, with owned storage capacities
and supported operations. Layouts map logical indices to physical storage; a
calling convention and runtime define buffers, transfers, and launches. Other
PIM architectures with these properties fit the same interface.

GPU and tensor accelerators also expose compute and storage hierarchies, but
leave loop order, multilevel tiling, spatial dataflow, allocation, routing,
instruction shape, and pipelining separately configurable
~\cite{chen2018tvm,parashar2019timeloop,kwon2019maestro}. PIM ownership instead
couples work mapping to layout: assigning an iteration to a unit constrains
where its operands can reside, and changing layout may require a new mapping or
explicit transfers. \Name therefore searches a constrained space of legal
mappings, layouts, operations, and transfers. Regular geometry makes layouts
algebraic (Section~\ref{sec:layout-motivation}); static ownership and explicit
transfers enable analytical ranking; and one mapping contract drives packing,
lowering, and native code generation. The host boundary is also architectural:
inputs may be broadcast, scattered,
or staged, while results may require gathering or reduction. A device-only
instruction count can therefore select a poor schedule. \Name records this
boundary in the target or runtime interface so modeled host transfers enter the
same cost graph as device operations.

\subsection{\texorpdfstring{Data Layout and the $\mathbb{F}_2$ Layout
Algebra}{Data Layout and the F2 Layout Algebra}}
\label{sec:layout-motivation}
\label{sec:background-ll}

\begin{figure*}[t]
  \centering
  \includegraphics[width=0.82\textwidth]{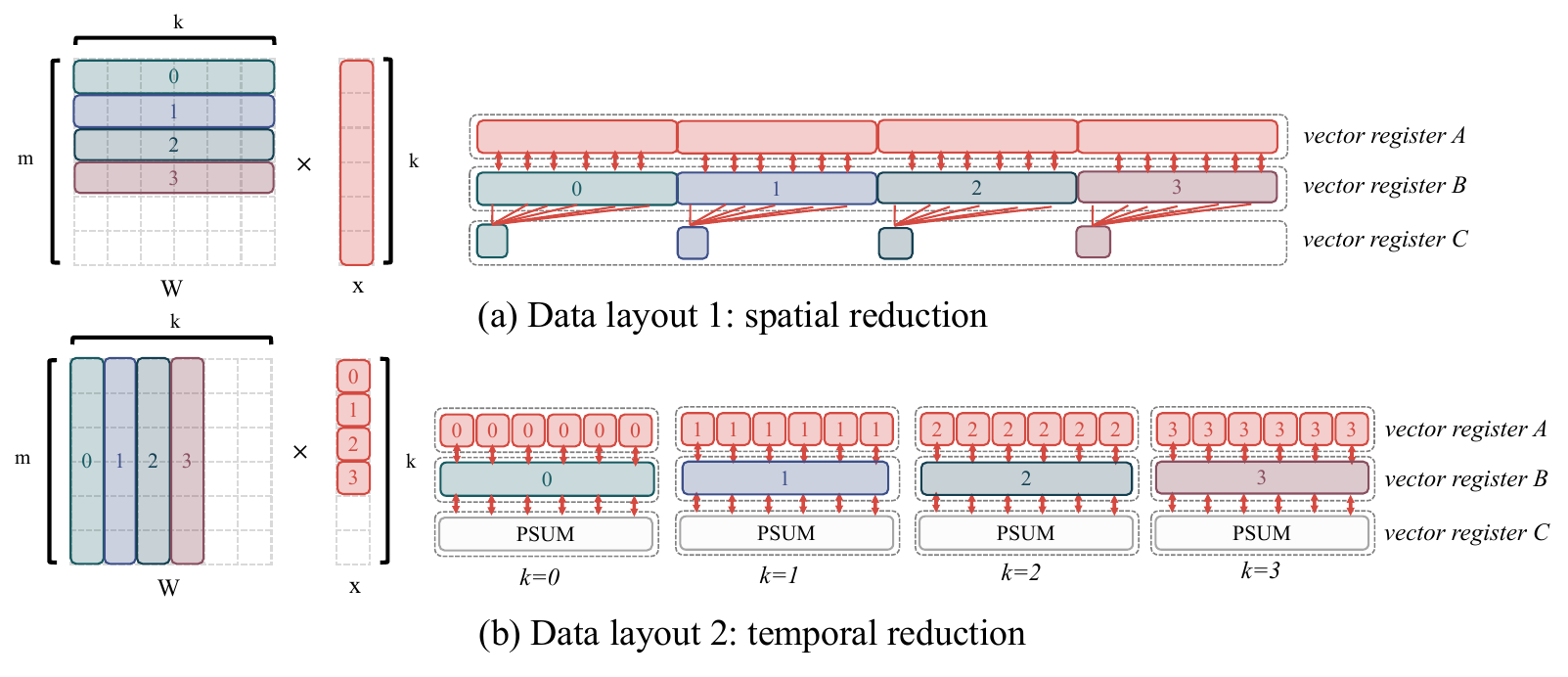}
  \caption{\textbf{APU data layouts trade reduction cost against lane
  utilization} -- For matrix-vector multiplication $y=Wx$ on GSI Gemini-I, a
  vector register holds 32{,}768 lanes, and a vector operation costs the same
  whether one lane or all of them are live.  (a) Cutting $W$ along $m$ puts $k$ in the
  lanes, so $x$ stays resident and the lanes fill for any $m$, but the partials
  for one output spread across lanes and \revise{must be summed across those lanes,
  costing about twenty times an elementwise add}.  (b) Cutting $W$ along
  $k$ keeps every output in one lane, so a broadcast $x[k]$ and an elementwise
  add advance the accumulator, at the price of one operand transfer per $k$
  step and $1-m/32768$ of each register left idle.  The group reductions in (a)
  scale with $m$ while the \revise{fraction of idle lanes} in (b) falls, so our cost program ranks
  (a) ahead by $128\times$ at $m{=}256$, by $11\times$ at $m{=}4096$, and by
  only $1.4\times$ once $m$ reaches the lane count; (b) wins outright when a
  second output axis supplies the parallelism.}
  \label{fig:apu-reduction-orientation}
  \Description{On the left of each panel, the product y = Wx with W drawn as a
  grid of blocks numbered 0 to 3 and x as a column.  On the right, three vector
  registers A, B, and C.  In panel (a) W is cut into row blocks, register A
  holds x replicated across lanes, register B holds one row block laid out
  k-contiguous, and converging lines collapse each k into a single
  scattered entry of register C.  In panel (b) W is cut into column blocks and
  four snapshots at k equals 0 through 3 show register A holding one broadcast
  element of x, register B holding one column of W, and register C accumulating
  a partial sum at the same lane.}
\end{figure*}

\noindent\textbf{Why layout decides PIM performance.}
On PIM hardware, data layout determines which resource serves an access, what
can remain resident, and which transfers the program performs.  Because a bank
also owns a compute unit, it decides more than access cost: the same coordinate
change can move an operand out of reach and turn an implicit read into an
explicit host round trip.  Prior
studies report order-of-magnitude swings from bank assignment, replication, and
position within a vector register~\cite{zhang2025apu,ibrahim2024pimnast,%
liu2025optipim}.  Figures~\ref{fig:samsung-gemv-layout}
and~\ref{fig:apu-reduction-orientation} provide concrete examples on Samsung
HBM-PIM and GSI Gemini-I, where changing the data layout changes operand
residency, reduction scope, and required movement.  Existing PIM compilers encode these decisions as
device-specific address arithmetic, so a layout that an expert discovers on one
target teaches the compiler nothing about the next.  What is missing is a
representation in which a layout is an object the compiler can construct,
compose, and check.

\revise{\noindent\textbf{Linear map representation.}
An axis of size $2^d$ needs $d$ bits to identify a position.
A linear layout~\cite{zhou2026linearlayouts} maps physical-coordinate bits
$u$ to logical tensor-index bits $t$ using a binary matrix $M$:
\[
  t=L(u)=Mu\pmod 2.
\]
This arithmetic is over $\mathrm{GF}(2)=\mathbb F_2=\{0,1\}$: addition is
XOR and multiplication is AND. Each output bit is the XOR of selected input
bits. Column $i$ of $M$ tells us which tensor-index bits change when input
bit $i$ flips. Bit permutations and XOR-based address rearrangements therefore
share one representation.}

\revise{\noindent\textbf{Distributed and shared memory layouts.}
On a GPU, a distributed layout maps a register index, thread index and warp
index to the tensor element stored there. Every tensor element appears at
least once; multiple locations may hold copies. A shared-memory layout maps
memory offsets to tensor elements, with address bits selecting a bank and a
position within that bank. GPU compilers use these maps to move values
between layouts and reduce shared-memory bank conflicts. PIM layouts use the
same representation with physical axes such as banks, rows, tasklets, vector
registers and lanes.}


\section{The \texorpdfstring{\Name}{Tenon} Language and Compiler}
\label{sec:abstraction}

\Name provides three small languages: a target language describes what a PIM
machine can do, a workload language describes what to compute, and a cost
language describes how expensive a generated implementation is.  They are
ordinary Python interfaces that build compiler \revise{intermediate representation (IR)}.
Table~\ref{tab:tenon-semantics} summarizes their roles.  This section defines
their semantics and illustrates each one with a GEMV example on Samsung HBM-PIM.

\begin{table}[t]
\centering
\footnotesize
\caption{\textbf{\Name language constructs} -- Target and workload grids
remain separate until compilation.}
\label{tab:tenon-semantics}
\renewcommand{\arraystretch}{1.08}
\begin{tabular}{@{}lp{0.72\columnwidth}@{}}
\toprule
\textbf{Construct} & \textbf{Semantics} \\
\midrule
\texttt{@tn.unit} &
Adds a physical axis; \revise{nesting states which unit contains another}. \\
\texttt{tn.mem}, \texttt{tn.reg} &
Adds storage owned by the current target unit. \\
\texttt{tn.op} &
Defines an operation's semantics, operands, and implementation. \\
\texttt{tn.move} &
Declares an explicit transfer between storage handles. \\
\texttt{@tn.work} &
Runs one body at each point in a logical work grid. \\
\texttt{@tn.rule} &
Maps a physical event to latency and resource use. \\
\texttt{tn.compile} &
Maps workload grids to target grids and ranks schedules. \\
\bottomrule
\end{tabular}
\end{table}

\subsection{Target Language}
\label{sec:abs-target}

A target is a nested grid of compute and storage units.
\code{\PYG{n+nd}{@tn}\PYG{o}{.}\PYG{n}{target}} creates the root, and each \code{\PYG{n+nd}{@tn}\PYG{o}{.}\PYG{n}{unit}} introduces a
named physical axis whose name and extent are given by \code{\PYG{n}{mapping}}.  Nesting
forms the machine hierarchy.  A memory or register declared inside a unit
belongs to that unit; its descendants may access it, while units outside that
subtree may not.  \revise{Thus, nesting in the target description states which units can access each memory.}
\code{\PYG{n+nd}{@tn}\PYG{o}{.}\PYG{n}{device}} names the device subtree; a unit with \code{\PYG{n}{mode}\PYG{o}{=}\PYG{l+s+s2}{\PYGZdq{}}\PYG{l+s+s2}{host}\PYG{l+s+s2}{\PYGZdq{}}}
names the host side of its transfers.

\code{\PYG{n}{tn}\PYG{o}{.}\PYG{n}{mem}} and \code{\PYG{n}{tn}\PYG{o}{.}\PYG{n}{reg}} declare storage and return \revise{handles---references to storage objects---that record
which unit contains the storage and its dimensions}.  \code{\PYG{n}{tn}\PYG{o}{.}\PYG{n}{op}} declares an operation over those
handles.  Its \code{\PYG{n}{fn}} gives the mathematical semantics used to match workload
expressions, while \code{\PYG{n}{emit}} gives the target implementation.  Its
\code{\PYG{n}{src}} and \code{\PYG{n}{dst}} arguments state where operands may reside, and
\code{\PYG{n}{tn}\PYG{o}{.}\PYG{n}{any\PYGZus{}}} denotes any member of a storage set.
\code{\PYG{n}{tn}\PYG{o}{.}\PYG{n}{move}} similarly declares a legal data path between two storage
handles.  Host collectives such as scatter, broadcast, and gather are moves,
not implicit runtime behavior.

\noindent\textbf{Samsung example.}
The following target excerpt declares 16 pseudo-channels, each containing
eight PIM units.  A unit selects its attached even/odd bank pair using its
symbolic coordinate and owns private general-purpose registers (GRFs).

\begin{Verbatim}[commandchars=\\\{\},fontsize=\scriptsize,baselinestretch=0.92,breaklines,
               linenos,xleftmargin=1.5em,numbersep=4pt,
               bgcolor=pimfill,bgcolorpadding=2pt]
\PYG{n+nd}{@tn}\PYG{o}{.}\PYG{n}{target}\PYG{p}{(}\PYG{l+s+s2}{\PYGZdq{}}\PYG{l+s+s2}{samsung\PYGZus{}hbm\PYGZus{}pim}\PYG{l+s+s2}{\PYGZdq{}}\PYG{p}{)}
\PYG{k}{def}\PYG{+w}{ }\PYG{n+nf}{samsung}\PYG{p}{(}\PYG{p}{)}\PYG{p}{:}
  \PYG{n+nd}{@tn}\PYG{o}{.}\PYG{n}{device}
  \PYG{k}{def}\PYG{+w}{ }\PYG{n+nf}{hbm}\PYG{p}{(}\PYG{p}{)}\PYG{p}{:}
    \PYG{n+nd}{@tn}\PYG{o}{.}\PYG{n}{unit}\PYG{p}{(}\PYG{n}{mapping}\PYG{o}{=}\PYG{p}{\PYGZob{}}\PYG{l+s+s2}{\PYGZdq{}}\PYG{l+s+s2}{channel}\PYG{l+s+s2}{\PYGZdq{}}\PYG{p}{:} \PYG{l+m+mi}{16}\PYG{p}{\PYGZcb{}}\PYG{p}{)}
    \PYG{k}{def}\PYG{+w}{ }\PYG{n+nf}{pseudo\PYGZus{}channel}\PYG{p}{(}\PYG{p}{)}\PYG{p}{:}
      \PYG{n}{banks} \PYG{o}{=} \PYG{n}{tn}\PYG{o}{.}\PYG{n}{mem}\PYG{p}{(}\PYG{n}{name}\PYG{o}{=}\PYG{l+s+s2}{\PYGZdq{}}\PYG{l+s+s2}{banks}\PYG{l+s+s2}{\PYGZdq{}}\PYG{p}{,} \PYG{n}{banks}\PYG{o}{=}\PYG{l+m+mi}{16}\PYG{p}{,} \PYG{o}{.}\PYG{o}{.}\PYG{o}{.}\PYG{p}{)}
      \PYG{n+nd}{@tn}\PYG{o}{.}\PYG{n}{unit}\PYG{p}{(}\PYG{n}{mapping}\PYG{o}{=}\PYG{p}{\PYGZob{}}\PYG{l+s+s2}{\PYGZdq{}}\PYG{l+s+s2}{pim}\PYG{l+s+s2}{\PYGZdq{}}\PYG{p}{:} \PYG{l+m+mi}{8}\PYG{p}{\PYGZcb{}}\PYG{p}{)}
      \PYG{k}{def}\PYG{+w}{ }\PYG{n+nf}{pim}\PYG{p}{(}\PYG{p}{)}\PYG{p}{:}
        \PYG{n}{\PYGZus{}}\PYG{p}{,} \PYG{n}{p} \PYG{o}{=} \PYG{n}{tn}\PYG{o}{.}\PYG{n}{get\PYGZus{}uid}\PYG{p}{(}\PYG{p}{)}
        \PYG{n}{even}\PYG{p}{,} \PYG{n}{odd} \PYG{o}{=} \PYG{n}{banks}\PYG{p}{[}\PYG{l+m+mi}{2}\PYG{o}{*}\PYG{n}{p}\PYG{p}{]}\PYG{p}{,} \PYG{n}{banks}\PYG{p}{[}\PYG{l+m+mi}{2}\PYG{o}{*}\PYG{n}{p}\PYG{o}{+}\PYG{l+m+mi}{1}\PYG{p}{]}
        \PYG{n}{a}\PYG{p}{,} \PYG{n}{acc} \PYG{o}{=} \PYG{n}{tn}\PYG{o}{.}\PYG{n}{reg}\PYG{p}{(}\PYG{o}{.}\PYG{o}{.}\PYG{o}{.}\PYG{p}{)}\PYG{p}{,} \PYG{n}{tn}\PYG{o}{.}\PYG{n}{reg}\PYG{p}{(}\PYG{o}{.}\PYG{o}{.}\PYG{o}{.}\PYG{p}{)}
        \PYG{n}{tn}\PYG{o}{.}\PYG{n}{move}\PYG{p}{(}\PYG{l+s+s2}{\PYGZdq{}}\PYG{l+s+s2}{LD\PYGZus{}A}\PYG{l+s+s2}{\PYGZdq{}}\PYG{p}{,} \PYG{n}{src}\PYG{o}{=}\PYG{n}{even}\PYG{p}{,} \PYG{n}{dst}\PYG{o}{=}\PYG{n}{a}\PYG{p}{,} \PYG{n}{emit}\PYG{o}{=}\PYG{o}{.}\PYG{o}{.}\PYG{o}{.}\PYG{p}{)}
        \PYG{n}{tn}\PYG{o}{.}\PYG{n}{op}\PYG{p}{(}\PYG{l+s+s2}{\PYGZdq{}}\PYG{l+s+s2}{MAC}\PYG{l+s+s2}{\PYGZdq{}}\PYG{p}{,} \PYG{n}{src}\PYG{o}{=}\PYG{p}{(}\PYG{n}{a}\PYG{p}{,} \PYG{n}{tn}\PYG{o}{.}\PYG{n}{any\PYGZus{}}\PYG{p}{(}\PYG{n}{banks}\PYG{p}{)}\PYG{p}{)}\PYG{p}{,} \PYG{n}{dst}\PYG{o}{=}\PYG{n}{acc}\PYG{p}{,}
              \PYG{n}{fn}\PYG{o}{=}\PYG{k}{lambda} \PYG{n}{x}\PYG{p}{,} \PYG{n}{w}\PYG{p}{,} \PYG{n}{acc}\PYG{p}{:} \PYG{n}{acc} \PYG{o}{+} \PYG{n}{x}\PYG{o}{*}\PYG{n}{w}\PYG{p}{,} \PYG{n}{emit}\PYG{o}{=}\PYG{o}{.}\PYG{o}{.}\PYG{o}{.}\PYG{p}{)}

  \PYG{n+nd}{@tn}\PYG{o}{.}\PYG{n}{unit}\PYG{p}{(}\PYG{n}{mode}\PYG{o}{=}\PYG{l+s+s2}{\PYGZdq{}}\PYG{l+s+s2}{host}\PYG{l+s+s2}{\PYGZdq{}}\PYG{p}{)}
  \PYG{k}{def}\PYG{+w}{ }\PYG{n+nf}{host}\PYG{p}{(}\PYG{p}{)}\PYG{p}{:}
    \PYG{n}{dram} \PYG{o}{=} \PYG{n}{tn}\PYG{o}{.}\PYG{n}{mem}\PYG{p}{(}\PYG{n}{name}\PYG{o}{=}\PYG{l+s+s2}{\PYGZdq{}}\PYG{l+s+s2}{host\PYGZus{}dram}\PYG{l+s+s2}{\PYGZdq{}}\PYG{p}{,} \PYG{o}{.}\PYG{o}{.}\PYG{o}{.}\PYG{p}{)}
    \PYG{n}{tn}\PYG{o}{.}\PYG{n}{move}\PYG{p}{(}\PYG{l+s+s2}{\PYGZdq{}}\PYG{l+s+s2}{SCATTER}\PYG{l+s+s2}{\PYGZdq{}}\PYG{p}{,} \PYG{n}{src}\PYG{o}{=}\PYG{n}{dram}\PYG{p}{,} \PYG{n}{dst}\PYG{o}{=}\PYG{n}{hbm}\PYG{o}{.}\PYG{n}{banks}\PYG{p}{,}
            \PYG{n}{verb}\PYG{o}{=}\PYG{n}{tn}\PYG{o}{.}\PYG{n}{scatter}\PYG{p}{,} \PYG{n}{emit}\PYG{o}{=}\PYG{o}{.}\PYG{o}{.}\PYG{o}{.}\PYG{p}{)}
\end{Verbatim}

The nesting states that a \revise{multiply-accumulate operation (MAC)} belongs to one PIM unit and can read only its
reachable banks and registers.  The \code{\PYG{n}{fn}} lets an ordinary multiply-add
in the workload match this instruction; the handles determine whether a
particular placement is legal.

\subsection{Workload Language}
\label{sec:abs-workload}

A workload is a grid of logical work items.  \code{\PYG{n+nd}{@tn}\PYG{o}{.}\PYG{n}{work}} defines a
grid whose shape is given by its \code{\PYG{n}{mapping}} argument, and every point
executes the same body.  \code{\PYG{n}{tn}\PYG{o}{.}\PYG{n}{get\PYGZus{}wid}\PYG{p}{(}\PYG{p}{)}} returns the current logical
coordinate, tensor arguments are shared across the grid, and local scalars are
private to the work item. The workload records typed operations, loops, tensor accesses, and
dependences.  

\noindent\textbf{Samsung example.}
For Samsung, we partition the output rows over a $16\times8$ logical grid.
Each work item owns 32 rows and expresses the reduction as ordinary tensor
code.

\begin{Verbatim}[commandchars=\\\{\},fontsize=\scriptsize,baselinestretch=0.92,breaklines,
               linenos,xleftmargin=1.5em,numbersep=4pt,
               bgcolor=pimfill,bgcolorpadding=2pt]
\PYG{n}{M}\PYG{p}{,} \PYG{n}{K}\PYG{p}{,} \PYG{n}{P}\PYG{p}{,} \PYG{n}{U} \PYG{o}{=} \PYG{l+m+mi}{4096}\PYG{p}{,} \PYG{l+m+mi}{1024}\PYG{p}{,} \PYG{l+m+mi}{16}\PYG{p}{,} \PYG{l+m+mi}{8}
\PYG{n}{ROWS} \PYG{o}{=} \PYG{n}{M} \PYG{o}{/}\PYG{o}{/} \PYG{p}{(}\PYG{n}{P} \PYG{o}{*} \PYG{n}{U}\PYG{p}{)}

\PYG{n+nd}{@tn}\PYG{o}{.}\PYG{n}{work}\PYG{p}{(}\PYG{n}{mapping}\PYG{o}{=}\PYG{p}{[}\PYG{n}{P}\PYG{p}{,} \PYG{n}{U}\PYG{p}{]}\PYG{p}{)}
\PYG{k}{def}\PYG{+w}{ }\PYG{n+nf}{gemv}\PYG{p}{(}\PYG{n}{W}\PYG{p}{:} \PYG{n}{f16}\PYG{p}{[}\PYG{n}{M}\PYG{p}{,} \PYG{n}{K}\PYG{p}{]}\PYG{p}{,} \PYG{n}{x}\PYG{p}{:} \PYG{n}{f16}\PYG{p}{[}\PYG{n}{K}\PYG{p}{]}\PYG{p}{,} \PYG{n}{y}\PYG{p}{:} \PYG{n}{f16}\PYG{p}{[}\PYG{n}{M}\PYG{p}{]}\PYG{p}{)}\PYG{p}{:}
  \PYG{n}{p}\PYG{p}{,} \PYG{n}{u} \PYG{o}{=} \PYG{n}{tn}\PYG{o}{.}\PYG{n}{get\PYGZus{}wid}\PYG{p}{(}\PYG{p}{)}
  \PYG{n}{row0} \PYG{o}{=} \PYG{p}{(}\PYG{n}{p} \PYG{o}{*} \PYG{n}{U} \PYG{o}{+} \PYG{n}{u}\PYG{p}{)} \PYG{o}{*} \PYG{n}{ROWS}
  \PYG{k}{for} \PYG{n}{r} \PYG{o+ow}{in} \PYG{n+nb}{range}\PYG{p}{(}\PYG{n}{ROWS}\PYG{p}{)}\PYG{p}{:}
    \PYG{n}{acc}\PYG{p}{:} \PYG{n}{f16} \PYG{o}{=} \PYG{l+m+mi}{0}
    \PYG{k}{for} \PYG{n}{k} \PYG{o+ow}{in} \PYG{n+nb}{range}\PYG{p}{(}\PYG{n}{K}\PYG{p}{)}\PYG{p}{:}
      \PYG{n}{acc} \PYG{o}{+}\PYG{o}{=} \PYG{n}{W}\PYG{p}{[}\PYG{n}{row0} \PYG{o}{+} \PYG{n}{r}\PYG{p}{,} \PYG{n}{k}\PYG{p}{]} \PYG{o}{*} \PYG{n}{x}\PYG{p}{[}\PYG{n}{k}\PYG{p}{]}
    \PYG{n}{y}\PYG{p}{[}\PYG{n}{row0} \PYG{o}{+} \PYG{n}{r}\PYG{p}{]} \PYG{o}{=} \PYG{n}{acc}
\end{Verbatim}

Here \code{\PYG{p}{(}\PYG{n}{p}\PYG{p}{,}\PYG{n}{u}\PYG{p}{)}} is a logical coordinate.  Mapping it to a pseudo-channel and
PIM unit is a schedule decision.  That schedule may \revise{divide the rows of $W$ among PIM units}, broadcast $x$, retain \code{\PYG{n}{acc}} in a local GRF, and gather $y$.

\begin{figure*}[t]
  \centering
  \includegraphics[width=\textwidth]{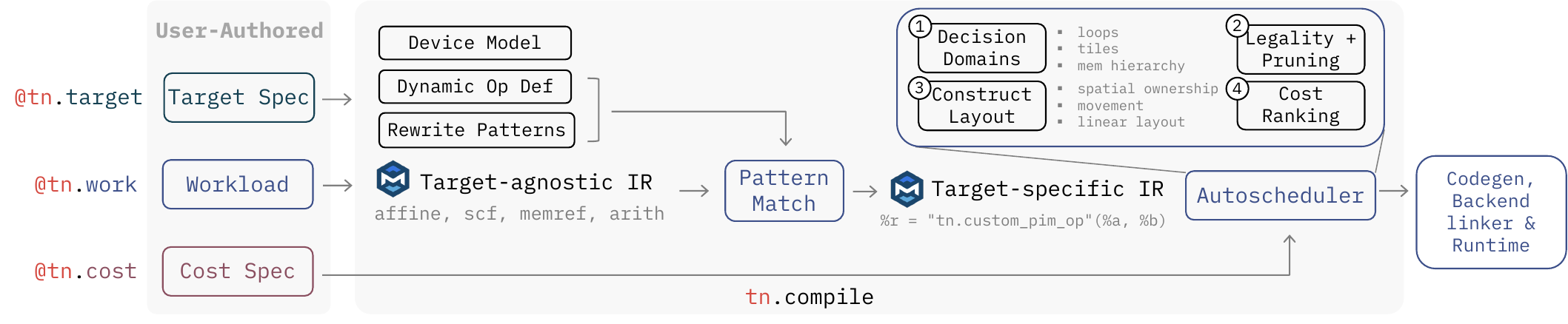}
  \caption{\textbf{\Name compilation flow} -- Workload and target
  specifications produce target-specific IR.  The autoscheduler then prunes
  candidates, constructs their layouts, and ranks them by cost before backend
  lowering.}
  \label{fig:tenon-compilation}
  \Description{The workload, target, and cost specifications feed Tenon's
  compiler.  The target produces a device model, dynamic operation
  definitions, and rewrite patterns.  A target-agnostic MLIR workload is
  pattern matched into target-specific IR.  The autoscheduler inset shows four
  numbered stages: decision domains, legality and pruning, layout construction,
  and cost ranking.  The selected program passes to backend code generation,
  linking, and runtime support.}
\end{figure*}

\subsection{Cost Language}
\label{sec:abs-cost}

\code{\PYG{n+nd}{@tn}\PYG{o}{.}\PYG{n}{cost}\PYG{p}{(}\PYG{n}{target}\PYG{o}{=}\PYG{n}{name}\PYG{p}{)}} defines a performance model separately from
target legality.  The model binds rules to target handles with
\code{\PYG{n+nd}{@tn}\PYG{o}{.}\PYG{n}{rule}\PYG{p}{(}\PYG{n}{handle}\PYG{p}{)}}.  Each rule receives a physical event and a context.
\code{\PYG{n}{ctx}\PYG{o}{.}\PYG{n}{step}} gives the event a latency, and \code{\PYG{n}{ctx}\PYG{o}{.}\PYG{n}{use}} reserves target
resources.  The evaluator schedules these activities subject to dependencies
and resource capacity, exposing both serialization and overlap.
\revise{Rule latencies come from whatever a backend offers: datasheet or
simulator timing for Samsung, AiM, AttAcc and UPMEM, and hardware
measurements for Gemini-I and Gemini-II. A coding agent can fit rules to device
microbenchmarks (Section~\ref{sec:eval-calibration}). A cost program captures
state-dependent timing, such as DRAM row hits, only if its rules encode it.}

\noindent\textbf{Samsung example.}
This rule prices the bank-to-GRF move declared above.  The move occupies its
PIM unit for 26 cycles but blocks the pseudo-channel issue resource for only
four.

\begin{Verbatim}[commandchars=\\\{\},fontsize=\scriptsize,baselinestretch=0.92,breaklines,
               linenos,xleftmargin=1.5em,numbersep=4pt,
               bgcolor=pimfill,bgcolorpadding=2pt]
\PYG{n+nd}{@tn}\PYG{o}{.}\PYG{n}{cost}\PYG{p}{(}\PYG{n}{target}\PYG{o}{=}\PYG{l+s+s2}{\PYGZdq{}}\PYG{l+s+s2}{samsung\PYGZus{}hbm\PYGZus{}pim}\PYG{l+s+s2}{\PYGZdq{}}\PYG{p}{)}
\PYG{k}{def}\PYG{+w}{ }\PYG{n+nf}{samsung\PYGZus{}cost}\PYG{p}{(}\PYG{n}{target}\PYG{p}{)}\PYG{p}{:}
  \PYG{n}{channel} \PYG{o}{=} \PYG{n}{target}\PYG{o}{.}\PYG{n}{unit}\PYG{p}{(}\PYG{l+s+s2}{\PYGZdq{}}\PYG{l+s+s2}{pseudo\PYGZus{}channel}\PYG{l+s+s2}{\PYGZdq{}}\PYG{p}{)}
  \PYG{n}{pim} \PYG{o}{=} \PYG{n}{target}\PYG{o}{.}\PYG{n}{unit}\PYG{p}{(}\PYG{l+s+s2}{\PYGZdq{}}\PYG{l+s+s2}{pim}\PYG{l+s+s2}{\PYGZdq{}}\PYG{p}{)}

  \PYG{n+nd}{@tn}\PYG{o}{.}\PYG{n}{rule}\PYG{p}{(}\PYG{n}{target}\PYG{o}{.}\PYG{n}{move}\PYG{p}{(}\PYG{l+s+s2}{\PYGZdq{}}\PYG{l+s+s2}{LD\PYGZus{}A}\PYG{l+s+s2}{\PYGZdq{}}\PYG{p}{)}\PYG{p}{)}
  \PYG{k}{def}\PYG{+w}{ }\PYG{n+nf}{load\PYGZus{}from\PYGZus{}bank}\PYG{p}{(}\PYG{n}{event}\PYG{p}{,} \PYG{n}{ctx}\PYG{p}{)}\PYG{p}{:}
    \PYG{n}{ctx}\PYG{o}{.}\PYG{n}{step}\PYG{p}{(}
      \PYG{n}{latency}\PYG{o}{=}\PYG{l+m+mi}{26}\PYG{p}{,}
      \PYG{n}{occupy}\PYG{o}{=}\PYG{p}{[}\PYG{n}{ctx}\PYG{o}{.}\PYG{n}{use}\PYG{p}{(}\PYG{n}{pim}\PYG{p}{,} \PYG{n}{cycles}\PYG{o}{=}\PYG{l+m+mi}{26}\PYG{p}{)}\PYG{p}{,}
              \PYG{n}{ctx}\PYG{o}{.}\PYG{n}{use}\PYG{p}{(}\PYG{n}{channel}\PYG{p}{,} \PYG{n}{cycles}\PYG{o}{=}\PYG{l+m+mi}{4}\PYG{p}{)}\PYG{p}{]}\PYG{p}{)}
\end{Verbatim}

Other rules price MACs, stores, and host transfers.  Replacing this cost
program may change which schedule wins, but cannot make an unsupported
operation or transfer legal.

\subsection{Composition}
\label{sec:abs-compose}
\code{\PYG{n}{tn}\PYG{o}{.}\PYG{n}{compile}} composes a workload, target, and cost program. The workload
and target define \revise{supported finite} legal schedules; each schedule maps logical work and
data to target units and storage, selects target operations, and inserts
declared transfers. \revise{Where the bounded search is used,} the compiler generates each candidate before the cost
program ranks it, so only legal, successfully generated programs can win.

\revise{Execution invokes the device runtime; \code{\PYG{n}{estimate}\PYG{p}{(}\PYG{p}{)}} requires an
cost model registered for that implementation. Combining partial sums in a
different order requires explicit permission.}

\subsection{Compilation Flow}
\label{sec:abs-flow}

Figure~\ref{fig:tenon-compilation} shows how the three languages become an
executable program.  The workload first lowers to target-agnostic IR in core
MLIR dialects, while the target specification produces a device model,
dynamic operation definitions, and rewrite patterns.  Pattern matching
replaces \revise{supported expression graphs} with target-specific operations while
preserving their loops, accesses, and dependences.  The autoscheduler then
maps operations and values, resolves transfers and the host--device interface,
and selects \revise{a program whose code and data transfers have been generated}.  Finally, the backend lowers the selected
IR to target commands or source code and links the required runtime; virtual
execution follows the same flow but returns only the cost result.

\revise{Within the supported layout search, ties in movement survive to
final cost ranking. Requiring the original floating-point accumulation order can rule out
splitting a reduction before search.
Backend code generators and runtime interfaces remain target-specific.}

\subsection{Autoscheduling}
\label{sec:abs-autoschedule}

Figure~\ref{fig:tenon-compilation} summarizes the shared autoscheduling
lifecycle: decision domains, legality and pruning, layout construction, and
cost ranking.  \Name does not impose one fixed schedule space on every PIM
target.  For each matched program, the target description exposes \revise{a finite set of choices for each
scheduling decision}.  Depending on the target, these cover tiling and parallelism,
placement and residency, reduction strategy, and data transfer paths and
granularity.  \revise{Choices} may depend on earlier assignments, so the compiler can
prune choices that violate dependences, capacity, or target capabilities
without \revise{enumerating every combination of choices}.  This dependency order is
a search optimization, not a greedy schedule: each surviving complete
assignment is evaluated as one program.

After pruning, the scheduler constructs the candidate layout and materializes
the exact target program, including buffers, transfers, generated code, and
host--device \revise{application binary interface (ABI), which specifies how the host passes buffers and arguments to the device}.  The final stage ranks only successfully materialized
candidates using \revise{the selected cost program} and deterministic tie breaking.
This ordering prevents an abstractly cheap choice
from winning when the target cannot emit or execute it.

The layout-construction stage uses linear layouts for spatial mapping within
the broader search.  For each structural candidate, it exposes locality,
replication, partial reductions, packing, and required boundary movement.
Temporal order, residency, and utilization remain whole-program scheduling
decisions priced by the cost program.  Section~\ref{sec:layout} next describes
how \Name constructs a minimum-movement layout for each candidate.


\section{Minimum-Movement PIM Layouts}
\label{sec:layout}

\revise{A linear layout maps physical storage indices to logical tensor indices
using a binary matrix (Section~\ref{sec:background-ll}). Some physical axes
of a PIM layout select a compute unit; we
call them \emph{compute axes}. The layout therefore decides which units need
copies of each input and which units hold partial results of each output. We first count this movement on a small example
(Section~\ref{sec:layout-example}), then generalize the count to any linear
layout and hierarchy of compute axes (Section~\ref{sec:layout-ownership}), and
finally search for a layout with minimum movement
(Section~\ref{sec:layout-minmove}).}

\revise{\subsection{Counting Movement in an Example}
\label{sec:layout-example}}

\begin{figure*}[t]
\centering
\includegraphics[width=0.9\textwidth]{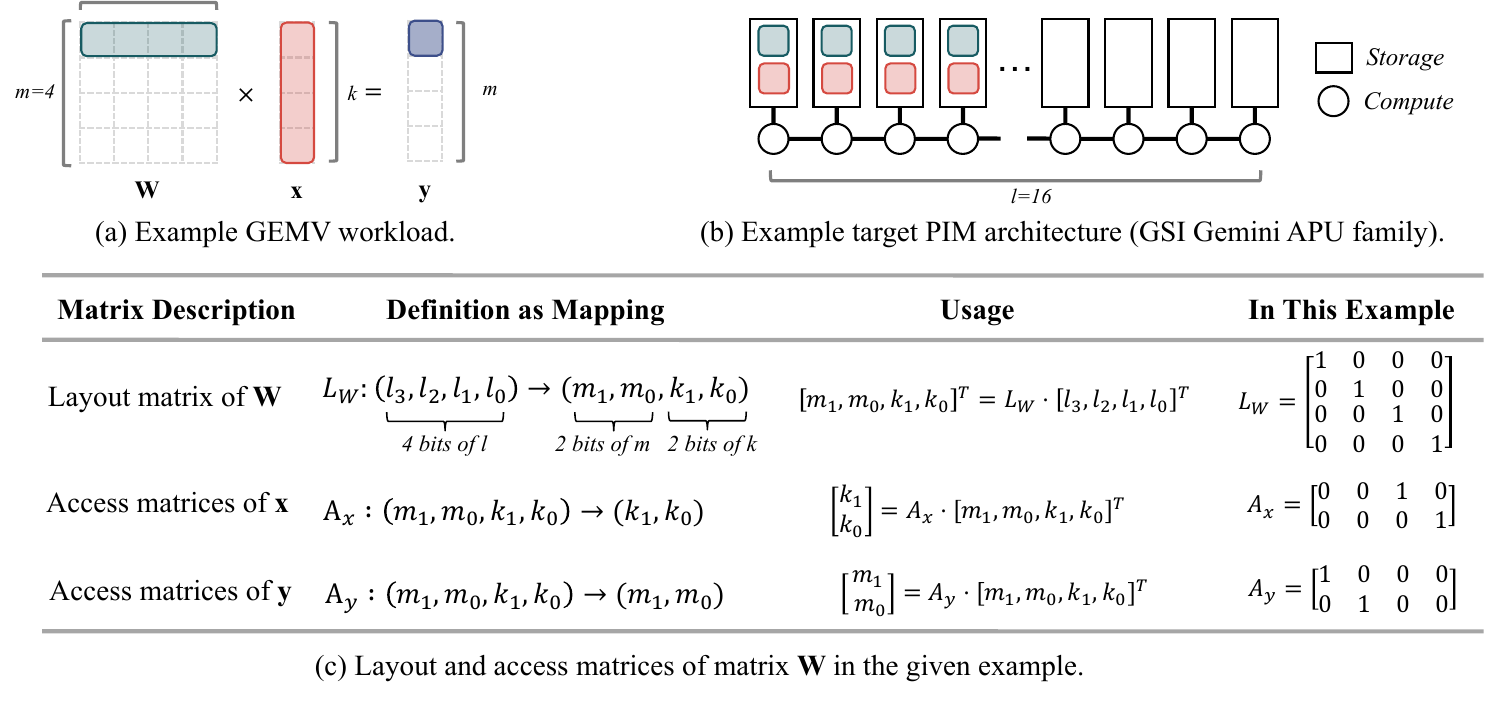}
\caption{\revise{\textbf{Example workload, target, and matrices.} (a) A $4\times4$ GEMV. (b) 16 lanes of a Gemini APU; lane $l=4m+k$ stores $W[m,k]$. (c) The layout matrix $L_W$ and access matrices $A_x$ and $A_y$.}}
\label{fig:ll-example}
\end{figure*}

\revise{Figure~\ref{fig:ll-example}(a,b) shows a GEMV $y=Wx$ with a $4\times4$
matrix on 16 lanes of a Gemini APU. Each lane pairs storage with a processing
element (PE), and neighboring PEs can exchange values. Loading consecutive
elements of $W$ and $x$ into consecutive storage units is cheap, whereas
moving data through the PE-to-PE connections is slow. The layout of $W$ maps a
lane, its storage index, to the weight that lane stores. We use the layout of
Figure~\ref{fig:apu-reduction-orientation}(a): lane $l=4m+k$ stores $W[m,k]$,
so each row of $W$ fills one group of four lanes, and each lane also holds
the input element its weight multiplies.}

\revise{\noindent\textbf{Layout and access matrices.}
We write a lane index as bits $(l_3,l_2,l_1,l_0)$ and a weight index as
$(m_1,m_0,k_1,k_0)$, where subscript 0 is the least significant bit.
Figure~\ref{fig:ll-example}(c) lists three matrices. The layout matrix $L_W$
maps a lane to the weight it stores; here it is the identity. The access
matrices come from the workload and do not depend on the layout: $A_x$
returns the index $k$ of the input a weight multiplies, and $A_y$ returns the
index $m$ of the output its product contributes to. Composing them with the
layout gives
\begin{equation}
  D_x=A_xL_W,\qquad D_y=A_yL_W,
  \label{eq:example-demands}
\end{equation}
which map each lane to the input it needs and the output it contributes to.
For example, lane 6 has bits $(0,1,1,0)$, stores $W[1,2]$, needs $x[2]$ and
contributes to $y[1]$.}

\begin{figure}[t]
\centering
\includegraphics[width=\columnwidth]{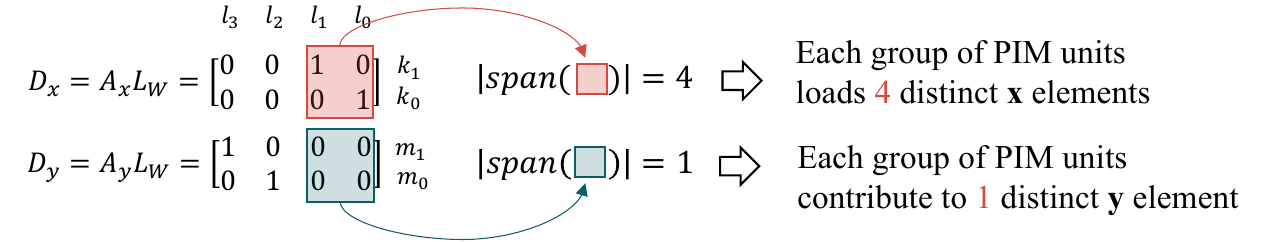}
\caption{\revise{\textbf{Counting the operands of one group of four lanes.} The XOR combinations of the shaded columns give the distinct elements.}}
\label{fig:ll-example-count}
\end{figure}

\revise{\noindent\textbf{Counting.}
To count the operands of one lane group, we fix the compute-axis bits that select the
group, $(l_3,l_2)$, and let the bits inside the group, $(l_1,l_0)$, vary.
Only the columns of $D_x$ and $D_y$ for the varying bits matter (shaded in
Figure~\ref{fig:ll-example-count}). The fixed bits add a constant offset, which
changes \emph{which} elements a group touches but not \emph{how many}. As
$(l_1,l_0)$ takes all four values, the group touches every XOR combination
of these columns, that is, their span. A set of columns with rank $r$, the
number of independent columns, spans $2^r$ elements. The shaded input
columns have rank 2 and the output columns rank 0, so each group needs 4
distinct elements of $x$ and contributes to 1 element of $y$. A zero block
spans $\{0\}$ and counts one element, not zero.}

\revise{These counts give the movement. All four groups need all four inputs,
so $x$ is replicated four times: 16 placements of 4 values, or 12 extra
copies. Each group contributes to one output element, but its four products
reside in separate lanes. Combining them requires moving partial sums through
the slower PE-to-PE connections within the group, even though no partial
result crosses groups. Assigning the bits the
other way, so that $(l_3,l_2)$ select $k$ and $(l_1,l_0)$ select $m$,
reverses the counts to one input and four outputs per group. Then no input is
replicated across groups, but each output has partial results in all four
groups that must be combined. This is the replication and reduction tradeoff
of Figures~\ref{fig:samsung-gemv-layout}
and~\ref{fig:apu-reduction-orientation}, read off two $2\times2$ blocks.
The cost program (Section~\ref{sec:abs-cost}) supplies the latencies of these
loads and PE-to-PE transfers, allowing the compiler to compare layouts using
the costs of the specific movements they require.}

\revise{\subsection{Generalizing Across Workloads and Targets}
\label{sec:layout-ownership}}

\revise{To generalize the example, we count how many compute units share
each tensor element. The same construction applies to workloads with any
number of iteration dimensions and tensors whose accesses are linear maps
of the iteration-index bits. For each input element, we identify the
iterations that read it; for each output element, those that contribute to
it. The layout maps these iterations to units, determining the replication
or reduction required at each level of the target hierarchy.}

\revise{Let $\mathcal I$ be the binary space of workload iteration indices.
We assume all coordinates are active; Section~\ref{sec:discussion} treats
masked domains and restrictions on reduction order. Let the target have $h$
nested compute levels, such as a lane group and a lane, or a DPU and a
tasklet. A layout is a one-to-one map
$L:\mathcal U_1\times\cdots\times\mathcal U_h\times\mathcal S
\rightarrow\mathcal I$, where $\mathcal U_s$ holds the compute-axis bits of
level $s$ and $\mathcal S$ the storage bits inside one unit. Let $Q_s$ keep
the compute-axis bits through level $s$ of $L^{-1}$, mapping an iteration to
its unit. For each tensor $j$, the access matrix $A_j$ maps an iteration
index to the tensor element it accesses. Iterations accessing the same
element differ by vectors in the kernel $\mathcal K_j=\ker A_j$: these
are changes to the iteration index that leave the tensor index unchanged.
Their units therefore differ by $Q_s(\mathcal K_j)$, so}
\begin{equation}
  r_{j,s}(L)
    =\operatorname{rank}\!\left(\left.Q_s\right|_{\mathcal K_j}\right)
  \label{eq:pim-movement-rank}
\end{equation}
\revise{gives the number of independent compute-index bits that vary, and $2^{r_{j,s}}$
units at level $s$ share each element of tensor $j$. This count applies
separately to every tensor at every compute level.}

\begin{table*}[t]
\centering
\caption{\textbf{Evaluation platforms and baselines} -- Platforms,
configurations, and metrics are backend-specific.}
\label{tab:eval-platforms}
\footnotesize
\renewcommand{\arraystretch}{1.08}
\begin{tabular}{@{}
  >{\raggedright\arraybackslash}p{0.11\textwidth}
  >{\raggedright\arraybackslash}p{0.17\textwidth}
  >{\raggedright\arraybackslash}p{0.17\textwidth}
  >{\raggedright\arraybackslash}p{0.27\textwidth}
  >{\raggedright\arraybackslash}p{0.18\textwidth}
  @{}}
\toprule
\textbf{Backend} & \textbf{Baseline} & \textbf{Execution Platform}
  & \textbf{Configuration} & \textbf{Reported Metric} \\
\midrule
Samsung HBM-PIM
  & \makecell[l]{Stock PIMSimulator\\GEMV}
  & PIMSimulator~\cite{saitpimsimulator}
  & \makecell[l]{64 independent pseudo-channels;\\16 banks and 8 PIM blocks\\
                 per pseudo-channel; FP16}
  & \makecell[l]{Placement, execution,\\and readback cycles} \\
\midrule
SK Hynix AiM
  & CENT Expert Traces
  & \makecell[l]{Ramulator2-based AiM\\simulator~\cite{luo2024ramulator2}}
  & \makecell[l]{32 GDDR6 channels;\\16 banks per channel}
  & Global memory-system cycles \\
\midrule
UPMEM
  & \makecell[l]{ATiM~\cite{shin2025atim},\\
    SimplePIM~\cite{chen2023simplepim},\\
    CINM~\cite{singh2024cinm}}
  & \makecell[l]{uPIMulator simulator\\\cite{hyun2024upimulator}}
  & \makecell[l]{1 DPU; 12 tasklets;\\350 MHz modeled clock}
  & Modeled DPU-kernel latency \\
\midrule
GSI Gemini-I
  & \makecell[l]{Expert-Optimized\\Phoenix Benchmark}
  & \makecell[l]{Physical Gemini-I\\Leda-E Card}
  & \makecell[l]{4 APUCs at 500 MHz;\\96 Mbit L1; 48 Mbit\\compute-in-memory SRAM}
  & \makecell[l]{\revise{Median device cycles}\\profiled by real-device\\execution} \\
\midrule
GSI Gemini-II~\cite{gsi2026gemini2matmul}
  & GTML Vendor Library
  & \makecell[l]{Physical Gemini-II\\Leda-E2 Card}
  & \makecell[l]{4 vector cores; 96 MiB L1;\\6 MiB associative compute SRAM;\\
                 PCIe Gen4$\times$16}
  & \makecell[l]{Device cycles\\profiled by real-device\\execution} \\
\midrule
AttAcc~\cite{park2024attacc}
  & DCC~\cite{yang2025dcc}
  & \makecell[l]{Ramulator2-AttAcc;\\analytical A100 roofline}
  & \makecell[l]{A100 + 5 HBM3 PIM devices;\\16 pCH/device; 64 banks/pCH; FP16}
  & \makecell[l]{PIM-kernel latency;\\GPT-13B decode speedup} \\
\bottomrule
\end{tabular}
\end{table*}

\revise{\noindent\textbf{Movement cost.}
Every unit beyond the first needs one more copy of an input or produces one
more partial result of an output. Let $B_j$ be the bytes of tensor $j$ and
$c_{j,s}$ the target's cost per byte of moving data across level $s$. The
movement cost of a layout is}
\begin{equation}
  C_{\mathrm{move}}(L)=
  \sum_j\sum_{s=1}^{h}
    c_{j,s}B_j
    \left(2^{r_{j,s}(L)}-1\right).
  \label{eq:pim-movement-cost}
\end{equation}
\revise{Inputs pay for replication and outputs pay for combining partial
results; transfers common to all layouts are omitted. Changing the workload
changes its access matrices and tensor sizes; changing the target changes
its compute hierarchy and movement costs. The same formulation therefore
covers different workloads and both host-mediated DRAM-PIM transfers and
movement inside an APU vector.}

\subsection{\revise{Finding the Minimum-Movement Layout}}
\label{sec:layout-minmove}

\revise{The compiler chooses}
\begin{equation}
  L^\star=\arg\min_{L\in\mathcal L_{\mathrm{legal}}}
             C_{\mathrm{move}}(L),
  \label{eq:pim-minmove-objective}
\end{equation}
\revise{where $\mathcal L_{\mathrm{legal}}$ contains the layouts that satisfy
the target's operation, capacity, alignment, and contiguity constraints.
Trying every layout is impractical: four bits already admit 20,160 one-to-one
layouts. Two techniques shrink the search.}

\revise{\noindent\textbf{Searching locality chains.}
The cost depends only on which iterations share a unit, not on where each
iteration sits inside its unit. Let $\mathcal T_s=\ker Q_s$ be the
directions that stay inside one unit through level $s$. Nested levels give a
chain $\mathcal I=\mathcal T_0\supseteq\mathcal T_1\supseteq\cdots\supseteq
\mathcal T_h$, and since
$r_{j,s}=\dim\mathcal K_j-\dim(\mathcal K_j\cap\mathcal T_s)$, the chain
fixes the cost. Here $\dim$ counts independent directions and $\cap$ keeps
directions common to two spaces. The search builds the chain one basis vector at a time,
starting from directions the operation forces to be local, and builds a full
layout only at the end. With four bits and three one-bit compute levels, the
20,160 layouts reduce to 315 chains.}

\revise{\noindent\textbf{Bounded pruning.}
For a partially built chain, assume that every remaining basis vector helps
every tensor. If $\mathcal T_s$ has $q$ of its $\ell_s$ basis vectors chosen,
spanning $\mathcal B$, this gives the optimistic rank}
\begin{equation}
  \underline r_{j,s}=\dim\mathcal K_j-
    \min\!\left(\dim\mathcal K_j,\,
       \dim(\mathcal B\cap\mathcal K_j)+\ell_s-q\right).
  \label{eq:pim-minmove-bound}
\end{equation}
\revise{Because the cost grows with each rank, substituting
$\underline r_{j,s}$ into Equation~\ref{eq:pim-movement-cost} never
overestimates any completion. Algorithm~\ref{alg:movement-search} expands the
branch with the lowest bound first and drops a branch only when its bound
strictly exceeds the best complete cost, so tied layouts remain for the cost
program to rank. On four bits with two or three compute levels, it visits 576 of
the 20,160 layouts.
In Algorithm~\ref{alg:movement-search}, $F$ stores partial layouts to explore,
$\mathrm{LB}(p)$ is the lower bound on movement cost for partial layout $p$,
$c^\star$ is the best complete cost, and $S$ contains layouts attaining it.}
\begin{algorithm}[t]
\revise{\caption{Minimum-movement search with all coordinates active}
\label{alg:movement-search}
\begin{algorithmic}[1]
\State $F\gets\{\text{required local assignments}\}$;
       $c^\star\gets+\infty$; $S\gets\varnothing$
\While{$F\ne\varnothing$ and budget remains}
  \State Remove a partial layout $p$ with lowest bound from $F$
  \If{$\mathrm{LB}(p)\le c^\star$}
    \If{$p$ is a complete legal layout}
      \State $c\gets C_{\mathrm{move}}(p)$
      \If{$c<c^\star$}
        \State $c^\star\gets c$; $S\gets\{p\}$
      \ElsIf{$c=c^\star$}
        \State $S\gets S\cup\{p\}$
      \EndIf
    \Else
      \State Insert legal extensions of $p$ into $F$ without duplicates
    \EndIf
  \EndIf
\EndWhile
\State \Return $(S,c^\star,\{\mathrm{LB}(p):p\in F\})$
\end{algorithmic}
}
\end{algorithm}

\revise{\noindent\textbf{Guarantee.}
A completed search returns every minimum-movement layout in the legal domain:
each legal layout determines a chain, which the search represents in a
fixed basis convention to avoid duplicates; the search can reach every chain,
and pruning never removes a minimum. A search that exhausts its budget
returns its best layout and the remaining lower bounds, not an optimality
claim. Least movement need not mean least runtime, so the cost program
makes the final choice among tied layouts (Section~\ref{sec:discussion}).}


\section{Evaluation}
\label{sec:evaluation}

\revise{We evaluate code performance across PIM targets, workload portability,
the effects of layout optimization, and the quality of cost-based selection.
An AttAcc study connects tiling decisions to modeled LLM decode latency.
We also measure programming and compilation cost and examine cost calibration
and virtual architecture exploration.}

\begin{figure*}[t]
  \centering
  \includegraphics[width=\textwidth]{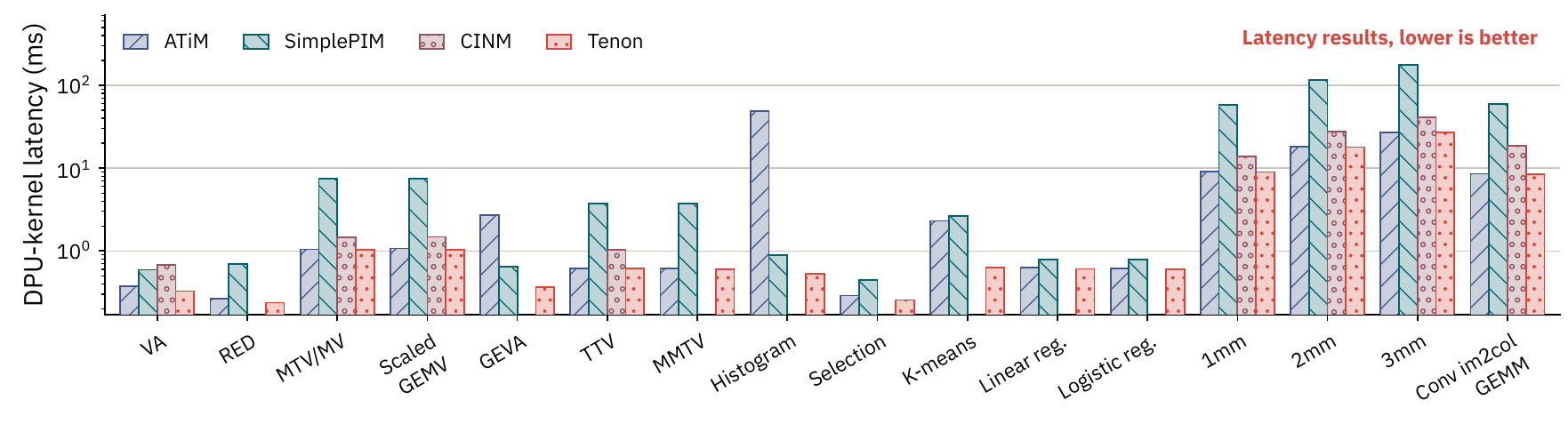}
  \vspace{-.3in}
  \caption{\textbf{UPMEM one-DPU kernel latency} -- uPIMulator with 12
  tasklets; log scale; lower is better. Missing bars denote unsupported
  kernels.}
  \label{fig:eval-upmem}
\end{figure*}

\subsection{Kernel Performance Across Targets}
\label{sec:eval-quality}

\subsubsection{Experimental Setup}
\label{sec:eval-platforms}

\revise{Table~\ref{tab:eval-platforms} lists the platforms, configurations and
native baselines. Gemini-I and Gemini-II use physical hardware; Samsung, AiM
and UPMEM use simulators. Speedup is baseline latency or cycles divided by
\Name's corresponding measurement. Workload suites and timing boundaries
differ across targets, so comparisons are within each target.}

\subsubsection{Samsung HBM-PIM}

\Name matches the vendor path exactly because its \texttt{GENERIC\_REDUCE}
lowering reconstructs the \revise{standard} GEMV command sequence
\cite{lee2021hbmpimstack,saitpimsimulator,saitpimlibrary}.
\revise{This comparison permits regrouping partial sums and includes placement,
execution and readback. Section~\ref{sec:eval-portability} examines execution
that instead preserves source-order FP16 arithmetic.}

\subsubsection{SK Hynix AiM}

\begin{table}[t]
\centering
\scriptsize
\caption{\textbf{SK Hynix AiM comparison with CENT} -- Global memory-system
cycles from the shared timing model; $L$ is the active sequence length.}
\label{tab:eval-skhynix}
\begin{adjustbox}{max width=\linewidth}
\begin{tabular}{lrrr}
\toprule
\textbf{Kernels} & \textbf{CENT (cycles)} & \textbf{\Name{} (cycles)}
  & \textbf{Speedup} \\
\midrule
RMSNorm + Residual, $L=128$       & 11,808  & 11,750  & $1.01\times$ \\
QKVO + RoPE, $L=128$              & 130,597 & 129,605 & $1.01\times$ \\
Attention QK + SV, $L=128$        & 14,272  & 11,010  & $1.30\times$ \\
Attention QK + SV, $L=512$        & 26,848  & 18,126  & $1.48\times$ \\
Attention QK + SV, $L=4096$       & 159,392 & 95,090  & $1.68\times$ \\
Softmax PIM Portion, $L=128$      & 6,307   & 6,307   & $1.00\times$ \\
Softmax PIM Portion, $L=512$      & 24,835  & 24,835  & $1.00\times$ \\
Softmax PIM Portion, $L=4096$     & 198,416 & 198,416 & $1.00\times$ \\
FFN Projections, $L=128$          & 250,433 & 250,433 & $1.00\times$ \\
FFN SiLU + Gating, $L=128$        & 11,775  & 11,775  & $1.00\times$ \\
Complete Mapped FFN, $L=128$      & 262,269 & 262,269 & $1.00\times$ \\
Mapped PIM Block, $L=128$         & 425,471 & 421,159 & $1.01\times$ \\
Mapped PIM Block, $L=512$         & 456,575 & 446,803 & $1.02\times$ \\
Mapped PIM Block, $L=4096$        & 762,668 & 697,316 & $1.09\times$ \\
\midrule
Geometric Mean                      &         &         & $1.10\times$ \\
\bottomrule
\end{tabular}
\end{adjustbox}
\vspace{-0.2in}
\end{table}

\revise{Compared with CENT's expert traces~\cite{gu2025cent}, \Name speeds up
attention by $1.30\times$ at context length 128 and $1.68\times$ at length
4,096 (Table~\ref{tab:eval-skhynix}). Softmax and feed-forward network (FFN)
latencies are equal in the two implementations, limiting the mapped PIM
block speedup to $1.09\times$ at length 4,096. The traces provide timing
comparisons but omit operand values, column addresses and accumulator
assignments; numerical validation of attention partitioning is evaluated
separately in Section~\ref{sec:eval-token-partition}.}

\subsubsection{UPMEM}

Across the 16 kernels in Fig.~\ref{fig:eval-upmem}, \Name is fastest on all 16,
with a geometric-mean $1.20\times$ speedup over the fastest supported baseline.
The gains concentrate in movement-heavy GEVA, histogram, and k-means
($1.70$--$3.65\times$), while dense matrix kernels remain within $2\%$ of the
fastest baseline. Coalesced layouts, buffered writes, and tiled reductions help
when they can remove MRAM transactions and increase WRAM reuse, rather than
change the arithmetic.

\revise{With host-side partitioning across 1--32 simulated DPUs (12 tasklets
each), fixed $391\times65$ GEMV latency falls from 1,149,430 to 38,484 cycles
on the slowest DPU, while fixed work per DPU stays at 73,803 cycles.
These device-only measurements exclude host transfer and merge costs; all 28
GEMV/reduction scaling cases pass numerical checks, including uneven partitions.}

\subsubsection{GSI Gemini-I}

\begin{table}[t]
\centering
\scriptsize
\caption{\textbf{Gemini-I comparison with expert Phoenix kernels} -- Median
device execution cycles on one physical \revise{APU core} and normalized workload
\revise{lines of code (LoC)}.}
\label{tab:eval-apuv1}
\setlength{\tabcolsep}{2.4pt}
\begin{adjustbox}{max width=\linewidth}
\begin{tabular}{lrrrrr}
\toprule
\textbf{Kernels}
  & \multicolumn{3}{c}{\textbf{Device Cycles}}
  & \multicolumn{2}{c}{\textbf{Workload LoC}} \\
\cmidrule(lr){2-4}\cmidrule(lr){5-6}
  & \textbf{Expert} & \textbf{\Name} & \textbf{Speedup}
  & \textbf{GVML} & \textbf{\Name} \\
\midrule
MatMul $256{\times}1024{\times}1024$
  & 8,784,431 & 8,764,900 & $1.002\times$ & 333 & 33 \\
K-Means $P32768,D24,K10$
  & 641,471 & 624,677 & $1.027\times$ & 263 & 46 \\
Histogram $N32768,B256$
  & 167,384 & 166,783 & $1.004\times$ & 141 & 33 \\
Linear Regression $8{\times}N4096$
  & 95,785 & 94,618.5 & $1.012\times$ & 175 & 38 \\
Word Count $Q128,N32768$
  & 175,397.5 & 171,988 & $1.020\times$ & 201 & 37 \\
\midrule
Geometric Mean
  & & & $1.013\times$ & & \\
\bottomrule
\end{tabular}
\end{adjustbox}
\end{table}

\revise{\Name reaches a geometric-mean $1.013\times$ speedup over expert
Phoenix kernels on physical Gemini-I (Table~\ref{tab:eval-apuv1})
\cite{zhang2025apu}, demonstrating expert-level performance for these kernels
and arithmetic rules. Section~\ref{sec:eval-layout-controls} isolates packing
effects from changes in operand residency.}

\subsubsection{GSI Gemini-II}

We report median \revise{cycles while each pipeline is running}, amortized over batches
of 20 with data transfers excluded.
The largest gain, $1.67\times$ on GESUMMV, occurs where two
matrix-vector products expose graph-level layout, fusion, and residency
choices; kernels closer to individual vendor calls gain
$1.07$--$1.12\times$. Across backends, \Name approaches parity when mapping
choices are less flexible and gains more when alternate layouts can eliminate
movement or preserve residency.

\subsection{Workload Portability and Numerical Correctness}
\label{sec:eval-portability}

\revise{We run the same $4096\times256$ FP16 GEMV source and rounding-sensitive
inputs on physical Gemini-I and Samsung PIMSimulator. Both match all 4,096
reference outputs when each multiply and addition rounds to FP16 in source
order. Regrouping additions changes results and therefore requires explicit
permission. Samsung's ordered fallback currently lacks cost modeling and layout
search. For irregular shapes, exhaustive checks of 121,968 small masked layouts
validate packing, reductions, active-unit costs and minima; masked search
disables bounds that assume all coordinates are active
(Section~\ref{sec:discussion}).}

\subsection{Layout Optimization and Search}
\label{sec:eval-layout}

\subsubsection{Shared Selection}

\revise{For one GEMV with reassociation permitted, shared analysis counts input
copies and partial outputs with equal weights across four Gemini-I tiles and
five DRAM partitions. Selected layouts drive packing and code generation.
The minimum-movement partition is fastest on DRAM, but Gemini-I tile 64 takes
$3.845\%$ longer than tile 256: movement pruning discards the faster candidate
before cost ranking. These are fixed candidate sets; DRAM timing uses DCC's
HBM3 model, excluding matrix preload and host reduction.}

\subsubsection{Packing Effects}
\label{sec:eval-layout-controls}

\revise{Across four GEMVs in the same HBM3 model and timing boundary, we compare
three packings with work assignment, registers, schedule, residency, bytes and
command counts fixed. A pre-timing row-transition rule selects layouts giving
$1.000$--$1.564\times$ speedup over row-major packing; the largest gain reduces
activations from 2,336 to 835. Outputs and addresses pass independent checks.
Two physical Gemini-I query-packing cases give a tie and $1.010\times$ across
seven interleaved rounds with computation, schedule and residency fixed.
The separate $12.71\times$ contraction gain also changes operand residency and
transfers.}

\subsubsection{Comparison with DCC}

\revise{Over 68 mappings across six GEMVs with common code generation, arithmetic
and timing, \Name wins once, ties four times and loses once against retrained
DCC ($1.007\times$ geometric-mean speedup). Each finds five measured minima.
Five DCC selections overlap training, limiting conclusions about unseen workloads.}

\subsubsection{Token Partitioning}
\label{sec:eval-token-partition}

\revise{\Name expresses PIMphony's token partitioning~\cite{kwon2026pimphony}
as channel placement with partial-result reduction. On eight simulated channels,
search over $1,2,4,8$ partitions reaches seven of nine measured minima
(maximum gap $2.63\%$), improving head-only placement by $1.071\times$ for one
head and 4,096 tokens. The nine workloads cross 1/8/32 heads with 128/512/4,096
context tokens at head dimension 128; all 45 variants pass numerical checks.
Controls include CENT-derived, head-only and fixed eight-way mappings.
Timing includes scoring, global softmax, value summation, host stages and
reductions, with a 32\,GB/s link and zero fixed latency. Inputs are resident;
cache append, projections and PIMphony's other optimizations are excluded.}

\subsection{Cost Modeling and LLM Decode on AttAcc}
\label{sec:eval-attacc}

\revise{We compare \Name's closed-form tiling selector with DCC's trained
XGBoost predictor~\cite{yang2025dcc}, then evaluate their impact on modeled
GPT-13B decode latency.}

\subsubsection{Tiling Quality and Prediction Error}

\begin{figure}[t]
  \centering
  \includegraphics[width=0.92\linewidth]{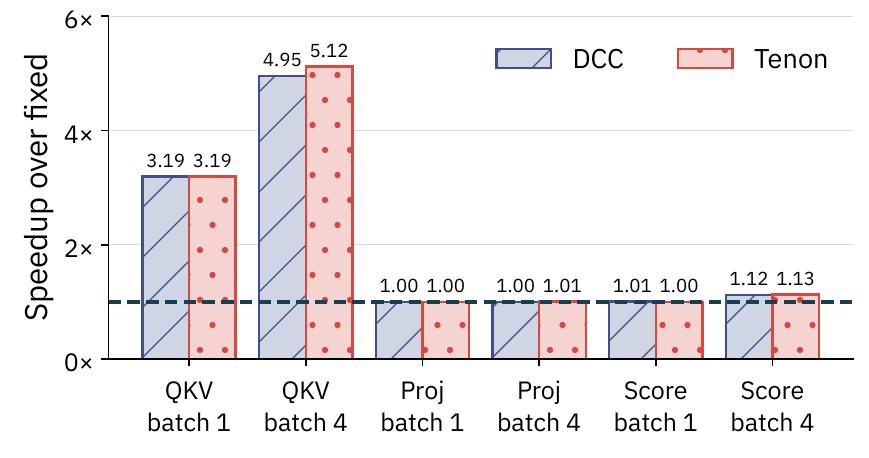}
  \caption{\textbf{AttAcc GPT-13B kernel tiling} -- Speedup over the fixed
  tiling for QKV, projection, and attention scoring at batches 1 and 4.
  Score bars report the mean over 248 decode-step shapes per batch; the
  dashed line marks the fixed tiling.}
  \label{fig:attacc-kernels}
  \vspace{-.2in}
\end{figure}

\revise{The 500 instances comprise query/key/value (QKV) and output-projection
GEMVs at batches 1 and 4, plus attention scoring at 248 shapes per batch.
Figure~\ref{fig:attacc-kernels} compares selected and fixed tilings.}

\revise{\Name matches the best reference simulator timing on 398 of 500
instances, compared with DCC's 326. Direct comparison with DCC gives 123 wins,
316 ties and 61 losses, all losses in attention. Among those losses, median
slowdown is $7.1\%$ relative to DCC and $9.4\%$ relative to the best reference
timing. The attention reference search stops using an assumed bound on
cost-model overprediction, so its best timings are not proven global minima.}

\revise{On 337 GEMV and 180 attention timing samples, the selection model's
Spearman rank correlations are 0.94 and 0.98, while mean absolute percentage
errors (MAPE) are $30.78\%$ and $14.19\%$. A separate command-replay model
captures more timing detail, reducing errors to $2.97\%$ and $6.30\%$ on the
same samples. It is not used for tiling selection.}

\revise{Replaying the worst and median losses shows ranking errors: faster
mappings are available, but the selected cost minimum is unique. In the worst
case, \Name takes 953 cycles versus 729; extra activations, precharges and
later final writes outweigh fewer MAC requests. The closed-form model omits
these row-state and issue-order effects.}

\subsubsection{Impact on Full-Model Decode}

\revise{GPT-13B decode latency combines Ramulator2-AttAcc PIM timings with
DCC's analytical GPU roofline, excluding prefill~\cite{park2024attacc}.
Both compilers place QKV/output projections, attention and softmax on PIM,
with other operations on GPU; only PIM tilings differ. FFN parallelism and
head-wise pipelining, including softmax, are modeled rather than measured
concurrent execution.}

\begin{figure}[t]
  \centering
  \includegraphics[width=\linewidth]{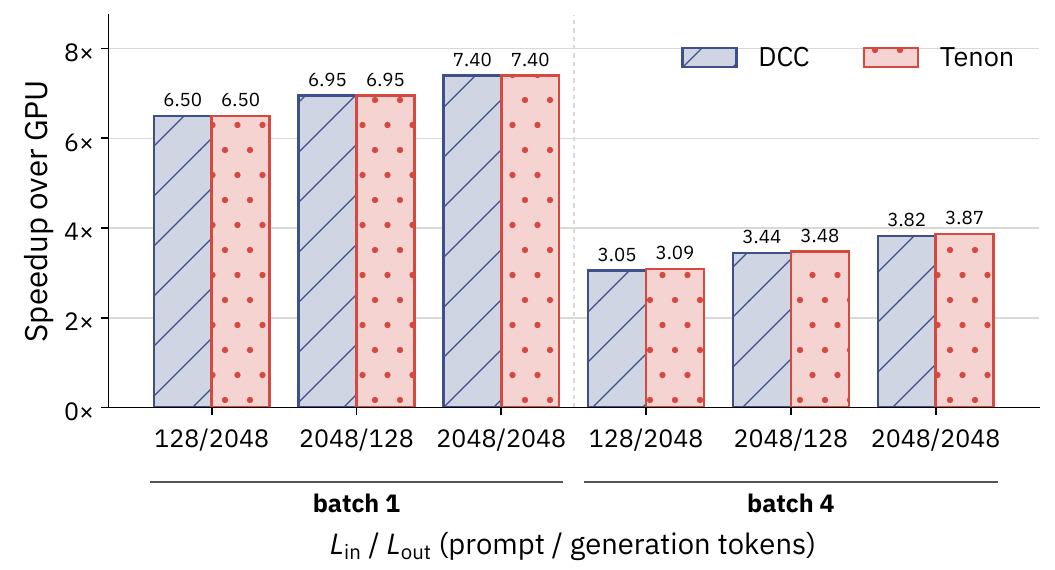}
  \caption{\textbf{AttAcc GPT-13B full-model decode} -- Speedup over the
  analytical GPU roofline for six prompt length, generation length, and
  batch-size configurations. Each x-axis label gives
  $L_\text{in}/L_\text{out}$.}
  \label{fig:attacc-e2e}
\end{figure}

\revise{Across the six configurations in Fig.~\ref{fig:attacc-e2e}, \Name and
DCC have equal speedups to two decimal places at batch 1, including one small
loss for \Name, while \Name reduces latency by $1.2$--$1.3\%$ at batch 4.
QKV explains most of the gain: at batch 4 with 128 input and 2,048 output
tokens, its latency falls from 1.2799 to 1.2332\,ms out of a total reduction
from 3.9933 to 3.9431\,ms. Equal FFN and other stage costs in both
compositions limit the application-level benefit.}

\subsection{Programming and Compilation Cost}
\label{sec:eval-attacc-compile}

\begin{table}[t]
\centering
\caption{\textbf{\Name--DCC workload LoC and compile time} -- \revise{Same candidate schedules and machine; Score is attention scoring, averaged over 248 sequence lengths.}}
\label{tab:attacc-compile}
\scriptsize
\setlength{\tabcolsep}{2.5pt}
\begin{tabular}{llrrrrr}
\toprule
Kernel & Batch & \revise{\#candidates}
  & \multicolumn{2}{c}{Workload LoC}
  & \multicolumn{2}{c}{Compile (ms)} \\
\cmidrule(lr){4-5}\cmidrule(lr){6-7}
& & & \Name & DCC & \Name & DCC \\
\midrule
\multirow{2}{*}{QKV (GEMV)}
  & 1 & 3{,}043 & \multirow{2}{*}{42} & \multirow{2}{*}{79} & 67 & 199 \\
  & 4 & 4{,}478 & & & 99 & 294 \\
\midrule
\multirow{2}{*}{Proj. (GEMV)}
  & 1 & 1{,}391 & \multirow{2}{*}{42} & \multirow{2}{*}{79} & 41 & 102 \\
  & 4 & 2{,}273 & & & 67 & 167 \\
\midrule
\multirow{2}{*}{Score (QK$^\top$)}
  & 1 & 946 & \multirow{2}{*}{28} & \multirow{2}{*}{126} & 36 & 77 \\
  & 4 & 1{,}704 & & & 59 & 133 \\
\midrule
\multicolumn{2}{l}{All 500 kernels}
  & 668{,}575 & 70 & 205 & 24\,s & 53\,s \\
\multicolumn{2}{l}{Offline sweep (per target)}
  & --- & --- & --- & \textbf{0} & $\sim$6\,h \\
\bottomrule
\end{tabular}
\end{table}

\revise{\Name reduces workload-description code and schedule-selection time
relative to DCC (Table~\ref{tab:attacc-compile}). For the 500 AttAcc instances,
it uses 66\% fewer workload lines and compiles $2.2\times$ faster on the same
machine and candidate schedules. Its analytical selector also avoids DCC's
six-hour predictor-training sweep.}

\revise{Each backend still requires a target-specific
runtime, which \Name neither generates nor shares across targets.}

\revise{\subsection{Agent-Assisted Cost Calibration}
\label{sec:eval-calibration}}

\revise{This proof-of-concept asks whether an executable cost program provides
a structured interface for agent-assisted calibration. In practice,
calibration is performed once per target, and the calibrated cost program is
reused across workloads.}

\begin{figure}[t]
  \centering
  \includegraphics[width=\columnwidth]{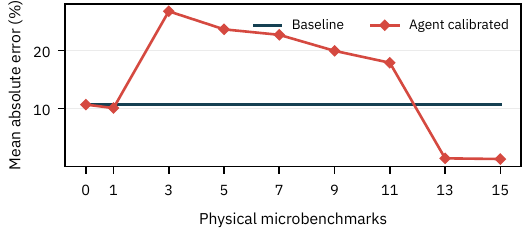}
  \caption{\textbf{Agent-assisted Gemini-I cost calibration} -- \revise{Mean
  absolute percentage error in full-kernel latency across three record-frequency
  configurations as component microbenchmarks are added. Baseline denotes the
  default, uncalibrated cost program.}}
  \label{fig:agent-calibration-learning-curve}
  \vspace{-.1in}
\end{figure}

\revise{Using the Codex harness with a GPT-5-family model, the agent fits these
costs from 15 physical component microbenchmarks. Composing the fitted costs
predicts three record-frequency kernel shapes not used for calibration and reduces mean
absolute latency error from $10.67\%$ to $1.29\%$
(Fig.~\ref{fig:agent-calibration-learning-curve}).}

\revise{We also reuse the fitted program on eight record-frequency plans and
eight histogram plans not used for calibration, with 64 physical launches per family and no
refitting. Mean error falls from $7.41\%$ to $1.23\%$ for record-frequency
and from $25.92\%$ to $0.41\%$ for histogram. For example, the fitted count
rule adds a 5,887.79-cycle start-up cost to 876.21 cycles per call.}

\subsection{Virtual-Backend Architecture Study}
\label{sec:eval-whatif}

\revise{A virtual backend compiles a workload for a target and returns a cost
estimate without invoking a device runtime (Section~\ref{sec:abs-flow}).
\Name's separate workload, target, and cost specifications let us explore
hardware that has not been built.
We vary target storage parameters while reusing the workload and cost rules:
the target determines legal tiles, and the cost program prices their generated
commands.}

\revise{For Llama2-7B's three FFN projections, we independently sweep AiM row
payload (1,024/2,048/4,096 BF16 elements) and Global Buffer capacity
(2/4/8\,KiB), fixing workload, DRAM capacity, channel/bank counts and timing
constants. Neither resource alone improves latency: both must accommodate a
larger reduction tile. Doubling both gives $1.30\times$ predicted and
$1.27\times$ simulated speedup; quadrupling gives $1.547\times$ and
$1.482\times$. Larger tiles reduce commands with diminishing benefit as
payload movement dominates. These unbuilt variants assume tiles can use a
prefix of a wider row, with buffer capacity enforced by target legality rather
than the simulator. Gains are conditional on these architectural and timing
assumptions.}


\section{Discussion and Limitations}
\label{sec:discussion}

\textbf{Analytical models still require validation.}
\revise{AttAcc choices are compared with the best reference Ramulator2-AttAcc
timings; the attention reference search uses an assumed model-error bound
to stop early. Gemini-I validates physical
point predictions on two workload families, with least squares on the same measurements giving
the same fit; these measurements do not establish broader schedule ranking.} The AiM study assumes fixed command timing and does not
establish silicon performance, area, power, or timing closure.

\revise{\textbf{Irregular shapes and floating-point order.}
Non-power-of-two axes are padded to a power-of-two size. A mask identifies
the valid coordinates;
packing, active units, and reductions respect that mask.
Equation~\ref{eq:pim-movement-cost} assumes every coordinate is active, so
padding alone does not make it a valid cost or lower bound. For an active set
$\mathcal D\subseteq\mathcal I$, let
$O_{j,s}(e)=\{Q_s(i):i\in\mathcal D,\ A_j(i)=e\}$ be the active units of
element $e$. For tensor $j$ at level $s$, the exact cost is
$c_{j,s}b_j\sum_{e\in A_j(\mathcal D)}(|O_{j,s}(e)|-1)$, where $b_j$
is the element size in bytes and $|O_{j,s}(e)|$ counts the units that use it.
Three active units therefore require two extra copies. Our masked search counts
these unit sets directly, disables the rank bound, and is exact only when it
completes. Permitted reductions also depend on the required floating-point order.
Splitting a floating-point reduction across units changes its rounding order
and requires explicit permission to regroup partial sums; a strict source-order
reduction must keep that order. Tests using exactly representable integer inputs do not prove equivalence for arbitrary floating-point inputs.
Movement optimization applies only after this constraint is satisfied.}

\textbf{Portability has a bounded scope.}
Targets share the workload/target/cost interface and \revise{selected search
infrastructure}, not schedule code,
ABIs, or runtimes; new formats still require \revise{code generators and launch code}. Each
backend searches a finite implemented space, and the prototype lacks general
allocation and \revise{spilling (moving values out of registers when they do not fit)}. Unsupported operations require a device or host
fallback. \revise{Weighted movement can also prune a faster runtime layout.}


\section{Related Work}
\label{sec:related}

\textbf{PIM compilers.}
CINM~\cite{singh2024cinm} provides MLIR infrastructure for heterogeneous
compute-in-memory and compute-near-memory devices, \revise{including shared tiling
and partial-result merging}, but its device dialects and
lowering paths remain target-specific.
\revise{DCC~\cite{yang2025dcc} co-optimizes data rearrangement and compute tiling
across a host/group/core hierarchy using a learned predictor.}
PIMFlow~\cite{kim2023pimflow} uses performance estimates to partition DNNs
\revise{between a GPU and Newton-based DRAM-PIM~\cite{he2020newton}}.  PrIM~\cite{gomez2022prim}, SimplePIM~\cite{chen2023simplepim},
and ATiM~\cite{shin2025atim} target UPMEM.
\revise{OptiPIM~\cite{liu2025optipim} derives hierarchical placement and data
layout from loop-bound choices, optimizing modeled execution cost under
processing-unit and storage constraints with integer linear programming (ILP).}
\Name separates workload semantics, target structure, and replaceable cost across
DRAM-PIM and SRAM-CIM while retaining target-specific lowering.

\revise{\textbf{Linear layouts.}
Linear Layouts~\cite{zhou2026linearlayouts} provides a binary-matrix
representation of data layouts and linear-algebra operations for manipulating
them. It is applied in GPU programming to reduce shared-memory bank conflicts.}

\textbf{Target-description systems.}
Exo 2~\cite{ikarashi2024exo2} defines hardware instructions and schedules;
TAIDL~\cite{mendis2024taidl} describes instruction semantics and synthesizes
compiler components; TVM~\cite{chen2018tvm} and VTA~\cite{moreau2019vta}
expose tensor instructions. \Name similarly separates semantics from emission,
but adds a nested PIM hierarchy and executable cost interface. Backend adapters
retain their input forms, code generators, and runtimes, and users may still
select explicit target primitives outside the automatic search.

\textbf{Analytical mapping and architecture models.}
Timeloop~\cite{parashar2019timeloop} and
MAESTRO~\cite{kwon2019maestro} search mappings using hierarchical
analytical models. \revise{\Name connects replaceable cost models to generated
programs, autoscheduler ranking and virtual backends; its interface also
supports roofline and learned models.}

\textbf{Scheduling and allocation.}
Halide~\cite{jrk2013halide}, TVM~\cite{chen2018tvm}, Spatial
~\cite{koeplinger2018spatial}, and Allo~\cite{chen2024allo} separate
computation from schedules. \Name specializes search to PIM hierarchy, host
staging, and heterogeneous memory tiers, generating each candidate before
estimating its cost.

\textbf{Accelerator generators.}
Gemmini~\cite{genc2021gemmini} and
Chipyard~\cite{amid2020chipyard} generate parameterized hardware. \Name instead
uses target-plus-cost programs for pre-silicon analysis and target-specific
backend adapters for compilation.


\section{Conclusion}
\label{sec:conclusion}

\Name shows that PIM backends can share a workload/target/cost interface,
layout optimization, and search without sharing an ISA or lowering path.
Across five commercial backends, \Name matches or outperforms vendor,
expert, and compiler baselines, with geometric-mean speedups up to
$1.20\times$ and individual speedups up to $3.65\times$. On AttAcc,
\Name-compiled GPT-13B workloads run $1.2$--$1.3\%$ faster than DCC across all
batch-4 configurations.

\begin{acks}
This work was supported in part by ACE, one of the seven centers in JUMP 2.0, a Semiconductor Research Corporation (SRC) program sponsored by DARPA, and by NSF under Award \#2403135.
\end{acks}


\clearpage
\bibliographystyle{IEEEtranS}
\bibliography{references}

\appendices
\section{AI use}

OpenAI Codex, using a GPT-5-family model, assisted with experimental-protocol
design, implementation and review of study code, artifact organization, data
visualization, and drafting or revision of
\revise{Sections~\ref{sec:eval-calibration} and~\ref{sec:eval-whatif}}.
The authors directed the research questions, reviewed and revised all generated
content, reran the experiments, and verified every reported numerical result
against archived raw logs and fingerprints. AI-generated output was not used
as a substitute for a physical or simulated measurement. The authors take
responsibility for the complete manuscript and artifacts.

\end{document}